\documentclass[aps,prd, twocolumn,groupedaddress, nobalancelastpage,nobibnotes, nofootinbib,longbibliography]{revtex4-1}
\pdfoutput=1
\usepackage{amsmath,amsfonts,amssymb,mathrsfs,graphicx,color}
\usepackage{verbatim}
\usepackage{enumerate}
\usepackage{graphicx}
\usepackage[hyperfootnotes=false]{hyperref}
\usepackage{xcolor}
\usepackage{slashed}
\usepackage[utf8]{inputenc}
\usepackage{chngcntr}
\usepackage{placeins}

\newcommand{\ie}{{\it i.e.}}

\newcommand{\eg}{{\it e.g.}}

\newcommand{\etc}{{\it etc.}}
\newcommand{\eq}{Eq.}

\newcommand{\fig}{Fig.}

\newcommand{\sibyll}[1]{{\sc Sibyll#1}}
\newcommand{\dpmjet}[1]{{\sc DPMJet#1}}
\newcommand{\qgsjet}{{\sc QGSJet}}
\newcommand{\eposlhc}{{\sc EPOS-LHC}}
\newcommand{\pythia}{{\sc pythia}}
\newcommand{\mceq}{{\sc MCEq}}
\newcommand{\rmd}{{\mathrm d}}
\newcommand{\xl}{x_{\rm Lab}}
\newcommand{\cm}{{c.m.\,}}

\newcommand{\equ}[1]{\eq~\eqref{eq:#1}}
\newcommand{\figu}[1]{\fig~\ref{fig:#1}}

\newcommand{\bi}{\begin{itemize}}
\newcommand{\ei}{\end{itemize}}

\newcommand{\revise}[1]{{\color{black}#1}}

\graphicspath{{plots/}{na49_plots/}{charm_plots/}{}}

\counterwithin{figure}{section}

\begin{document}


\title{The hadronic interaction model \sibyll{~2.3c} and inclusive lepton fluxes}

\author{Anatoli Fedynitch}
\affiliation{DESY, Platanenallee 6, 15738 Zeuthen, Germany}
\affiliation{Karlsruher Institut f\"ur Technologie, Institut f\"ur Kernphysik,
Postfach 3640, 76021 Karlsruhe, Germany}
\affiliation{Institute for Cosmic Ray Research, the University of Tokyo,
5-1-5 Kashiwa-no-ha, Kashiwa, Chiba 277-8582, Japan}

\author{Felix Riehn}
\affiliation{Laborat\'{o}rio de Instrumenta\c{c}\~{a}o e F\'{i}sica Experimental de Part\'{i}culas (LIP) - Lisbon, Av.\ Prof.\ Gama Pinto 2, 1649-003 Lisbon, Portugal}
\affiliation{Karlsruher Institut f\"ur Technologie, Institut f\"ur Kernphysik, Postfach 3640, 76021 Karlsruhe, Germany}

\author{Ralph Engel}
\affiliation{Karlsruher Institut f\"ur Technologie, Institut f\"ur Kernphysik, Postfach 3640, 76021 Karlsruhe, Germany}

\author{Thomas K. Gaisser}
\affiliation{Bartol Research Institute, Department of Physics and Astronomy, University of Delaware, Newark, DE 19716, USA}

\author{Todor Stanev}
\affiliation{Bartol Research Institute, Department of Physics and Astronomy, University of Delaware, Newark, DE 19716, USA}

\date{\today}

\begin{abstract} 
Muons and neutrinos from cosmic ray interactions in the atmosphere originate from decays of mesons in air-showers. \sibyll{-2.3c} aims to give a precise description of hadronic interactions in the relevant phase space for conventional and prompt leptons in light of new accelerator data, including that from the LHC. \sibyll{} is designed primarily as an event generator for use in simulation of extensive air showers.  Because it has been tuned for forward physics as well as the central region, it can also be used to calculate inclusive fluxes.  The purpose of this paper is to describe the use of \sibyll{-2.3c} for calculation of fluxes of atmospheric leptons.
\end{abstract}

\maketitle


\section{Introduction}
The main theme of this paper is the connection between hadronic interactions at high energies and the inclusive spectra of atmospheric leptons. The coupled transport equations that relate the lepton spectra to the primary cosmic-ray spectrum depend on the properties of the hadronic interactions, implemented here with \sibyll{-2.3c}.  A brief introduction to these transport equations is given in \ref{sec:cascade_physics}. The numerical methods are described in section \ref{sec:calc_method}. Section \ref{sec:had-prod} establishes the connection between particle physics observables and the regions of phase space that are important for inclusive lepton observables.  A key observation is that, because of the steep primary cosmic-ray spectrum, it is the forward fragmentation region of hadronic interactions that is of special importance for inclusive lepton spectra. In the second part of the paper (Section \ref{sec:sib23}) we describe how \sibyll{-2.3c} deals with the forward fragmentation region including production of charm. In Section~\ref{sec:impact} we summarize the impact of \sibyll{-2.3c} on inclusive lepton spectra.  We compare with the corresponding results of its predecessor, \sibyll{-2.1}, and with several other event generators in current use.  We try, as far as possible, to relate observed differences to specific features of hadronic interactions with the idea that precise measurements of atmospheric lepton spectra have the potential to constrain features of forward production of hadrons at high energy.

\section{Physics of atmospheric muons and neutrinos}
\label{sec:cascade_physics}
Cosmic rays interacting in the atmosphere produce secondary hadrons whose decay products give rise to a spectrum of atmospheric muons and neutrinos. The primary cosmic-ray energy spectrum is approximately a broken power-law and its nuclear mass composition varies as a function of energy. As a special form of the Boltzmann transport equations, the coupled cascade equations describe the evolution of particle fluxes in a dense or gaseous medium. The average number of interactions of a particle with air nuclei is a function of the slant depth or grammage
\begin{equation}
\label{eq:slant_depth}
  X(h_0) = \int_0^{h_0}~{\rm d}\ell~\rho_\text{air}(\ell),
\end{equation}
where $\rho$ is the density of the atmosphere and $h_0$ the altitude above the surface. At high energies inelastic hadronic interactions dominate and result in secondary particle production. Stable and longer lived particles can again interact and produce sub-cascades similar to the initial one but at reduced energy. Unstable particles decay into other hadrons or leptons or re-interact, depending on their energy and lifetime. The cascade evolution stops as the hadrons fall below the threshold for inelastic interactions, leaving only stable hadrons and atmospheric leptons in the cascade. The number of nucleons and mesons increases up to a certain depth or altitude and then attenuates. The production of muons and neutrinos is proportional to the number of decaying mesons.  Lepton production therefore decreases at lower altitudes as the atmospheric density increases.
For the same reason, the flux of leptons from decay of pions and kaons increases as
zenith angle increases. At large inclinations, low energy muons decay in flight, but higher-energy muons can survive and reach the ground.

\subsection{Coupled cascade equations}
The equations describe the evolution of the differential flux, defined as the differential of the particle
flux $\phi$ with respect to energy per unit area, unit solid angle, and time
\begin{equation}
  \Phi = \frac{\rmd \phi}{\rmd E} = \frac{\rmd N}{\rmd E\ \rmd A\ \rmd \Omega\ \rmd t}.
\end{equation}
The coupled cascade equations
\begin{equation}
\begin{split}
\frac{\rmd \Phi_h(E,X)}{\rmd X}  = 
 &-\frac{\Phi_h(E,X)}{\lambda_{\text{int}, h}(E)}\\
 &-\frac{\Phi_h(E,X)}{\lambda_{\text{dec},h}(E, X)}\\
 &-\frac{\partial}{\partial E}(\mu(E) \Phi_h(E,X))\\
 &+ \sum_\ell{\int_E^\infty \rmd E_\ell~ \frac{\rmd N_{\ell(E_\ell) \to h(E)}}{\rmd E} 
                          \frac{\Phi_\ell(E_\ell,X)}{\lambda_{\text{int}, l}(E_\ell)}}\\
 &+ \sum_\ell{\int_E^\infty \rmd E_\ell~ \frac{\rmd N^\text{\text{dec}}_{\ell(E_\ell) \to h(E)}}{\rmd E} 
                          \frac{\Phi_\ell(E_\ell,X)}{\lambda_{\text{dec}, l}(E_\ell, X)}},
\end{split}
\label{eq:cascade_equations}
\end{equation}
are the transport equations representing the interplay between the two dominating high energy processes, interactions and decays. The energy losses \revise{($\mu(E)$)} from ionization and multiple scattering ($\langle\rmd E/\rmd X\rangle \approx 2\, \text{MeV}\,\text{cm}^2/\text{g}$) 
 of muons impact muon and electron neutrino fluxes below tens of GeV for near-vertical or below few TeV for near-horizontal directions.  
\begin{figure}
  \includegraphics[width=.95\columnwidth]{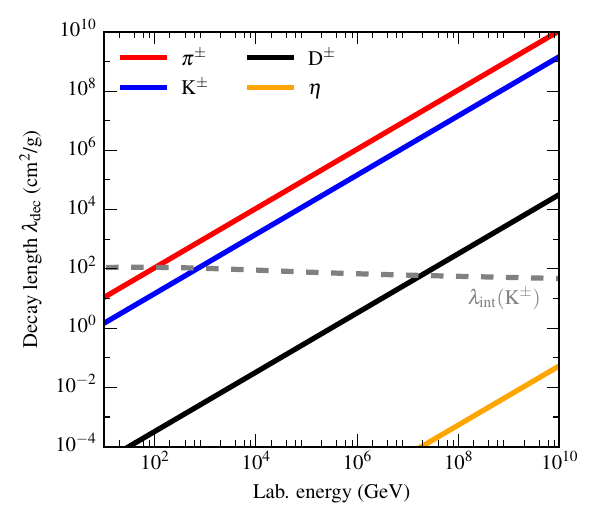}
  \caption{\label{fig:int_dec_len} Decay lengths for relevant mesons ({\rm solid}) compared to the interaction length for kaons calculated at $8\,$km altitude a.s.l.}
\end{figure}
The energy dependence of the interaction lengths
\begin{equation}
\label{eq:int_length}
  \lambda_{\text{int},h}(E) = \frac{\langle m_\text{air} \rangle}{\sigma^{\rm inel}_{h-\text{air}}(E)}.
\end{equation}
and decay lengths
\begin{equation}
\label{eq:dec_length}
\lambda_{\text{dec},h}(E, X) = \frac{c\tau_h E \rho_\text{air}(X)}{m_h c^2}
\end{equation}
are shown in \figu{int_dec_len}. For short-lived particles $\lambda_{\text{dec}} \ll \lambda_{\text{int}}$, so the secondaries preserve the energy spectrum of their mother species. At the other extreme, where hadronic interactions dominate, the flux of \revise{leptons} is attenuated and becomes a power steeper \cite{Gaisser:2016uoy}. The transition between these two regimes (cross-over between solid and dashed lines in \figu{int_dec_len}) is called critical energy $\varepsilon_h$ and is a function of the density or the altitude. For typical conditions the values are approximately $115\,$GeV for $\pi^\pm$, $850\,$GeV for $\text{K}^\pm$, $10\,$PeV for $\text{D}$-mesons and $> 10^{13}\,$GeV for unflavored $\eta$ mesons. 

\revise{The general features of the atmospheric lepton spectrum can be understood from approximate} semi-analytical solutions (see \revise{for more details} \cite{Gaisser:2016uoy,Lipari:1993hd,Fedynitch:2012fs}) of the form
\begin{equation}
\label{eq:semi-anal}
\Phi_\ell(E) = \frac{\phi_\text{N}(E)}{1 - Z_{\rm NN}}\sum_{\substack{h = \pi,\\ \text{K}, \text{K}^0_\text{L},\dots}} \frac{Z_{\text{N}h,\gamma} Z_{h \to \ell,\gamma}}{1 + B_h E \cos\theta/\varepsilon_h}.
\end{equation}
The energy dependence of the lepton flux follows the cosmic ray nucleon flux and becomes a power steeper above the critical energy, re-scaled by the cosine of the zenith angle $\theta$. The $Z$ factors are (primary cosmic ray) spectrum weighted moments, representing hadronic interactions and particle decays
\begin{equation}
\label{eq:z-factor}
  Z_{Nh} = \int_0^1 \rmd{} \xl ~ \xl^{\gamma - 1} \frac{\rmd N_{\text{N} \to h}}{\rmd \xl}.
\end{equation}
The values of the spectral index $\gamma$ are between 2.7 and 3.0 (see Sect.\ \ref{sec:primary_flux_H3a}). The energy fraction is defined as $\xl = E_\text{secondary}/E_\text{projectile}$. The weight emphasizes the very forward part of the particle spectrum with $\xl \gtrsim 0.2$, or in other words, particles with very small scattering angles. The standard approximation is scaling, \ie{}\ the secondary particle spectrum is only a function of $\xl$ and independent of the nucleon's energy. In practice and in current measurements (\figu{pi0-scaling}) this approximation is valid in the very forward phase space. But at smaller $\xl$ and high energies scaling is known to be violated due to multiple parton interactions in one collision.

For very short lived hadrons, such as charmed or unflavored mesons the second term in the denominator of \equ{semi-anal} is negligible. This particular fraction of the lepton flux thus follows the spectral index of the primary spectrum up to very high energies. Due to the attenuation-less immediate decay of these mesons, the resulting lepton flux is called prompt. The number of prompt muons and neutrinos is small because the mother mesons are rarely produced (small $Z_{\text{N},h,\gamma}$) and because the leptonic decay branching ratios are small ($Z_{h \to \ell,\gamma}$). All other inclusive leptons (mainly from decay of pions and kaons) constitute the conventional component, which is suppressed at very high energies $> 100\,$TeV so that the small prompt flux eventually dominates the total rate of atmospheric muons and neutrinos.

\section{Calculation method}
\label{sec:calc_method}
We carry out the computation of inclusive atmospheric lepton fluxes with a numerical method described in detail in \cite{cpc_paper,Fedynitch:2015zbe,Fedynitch:icrc2013,Fedynitch:2015zma,Fedynitch:2016nup}. This method is more powerful and precise than the semi-analytical solutions, especially if the cosmic-ray flux is not a single power law or when scaling violations have to be taken into account. Monte-Carlo calculations \cite{Fedynitch:2012fs,Honda:2015fha} achieve comparable accuracy but they become inefficient at energies above several hundreds GeV. Biasing techniques can reduce this inefficiency that is related to the absorption of mesons in the atmosphere, however, no such technique is available at present to bias hadronic interaction models for the generation of mesons carrying a large fraction of the projectiles momentum.

\subsection{Matrix cascade equations}
\revise{A numerical solver for the system of the discretized coupled cascade equations}
\begin{equation}
\label{eq:cascade_equations_dicr}
\begin{split}
\frac{\rmd \Phi^h_{E_i}}{\rmd X}  =  &-\vec{\nabla}_i(\mu^h_{E_i}\Phi^h_{E_i}) \\
 &-\frac{\Phi^h_{E_i}}{\lambda^h_{\text{int}, E_i}}  + 
 \sum_{E_k \ge E_i}^{E_N}
 {\sum_\ell{\frac{c_{\ell(E_k) \to h(E_i)}}{\lambda^\ell_{\text{int},E_k}}\Phi^\ell_{E_k}}}\\
 &-\frac{\Phi^h_{E_i}}{\lambda_{\text{dec},E_i}^h(X)}
 + \sum_{E_k \ge E_i}^{E_N}
 {\sum_\ell{\frac{\mathrm{d}_{\ell(E_k) \to h(E_i)}}{\lambda^\ell_{\text{dec},E_k}(X)}\Phi^\ell_{E_k}}}.
\end{split}
\end{equation}
\revise{is implemented in the open-source software Matrix Cascade Equations (\mceq)\footnote{\url{https://github.com/afedynitch/MCEq}} \cite{Fedynitch:2015zma}}. The index $h$ represents one of the $\sim 65$ particle species and the energy index $i$ runs over an energy grid, subdivided into 10 logarithmically spaced bins per decade of energy across 14 orders of magnitude ($100\,$MeV - $1\,$ZeV). The coefficients $c_{\ell(E_k) \to h(E_i)}$ represent inclusive secondary particle energy spectra in the target laboratory frame, and $d_{\ell(E_k) \to h(E_i)}$ the corresponding decay spectra ${\rm d}N^{(dec)}_{l(E_l) \to h(E)}/\mathrm{d}E$ from \equ{cascade_equations}. The solver performs a simultaneous integration of a coupled system of $N_{\rm species} \times N_{\rm E-bins} \sim 8000$ ordinary differential equations (ODEs) in parallel. For particles which suffer continuous energy losses, the ODEs become partial differential equations. This Fokker-Planck part of equation is solved through a finite differences operator of order-3 or higher. The resulting expression in matrix form is
\begin{equation}
\begin{split}
  \frac{\rm{d}}{\mathrm{d}X}\vec{\Phi} = -\vec{\nabla}_E(\text{diag}({\vec{\mu}}) \vec{\Phi})
                                          + (-\mathbf{1} + \mathbf{C}){\mathbf{\Lambda}}_\text{int}\vec{\Phi}&  \\
                                          + \frac{1}{\rho(X)}(-\mathbf{1} + \mathbf{D})\mathbf{\Lambda}_\text{\text{dec}}\vec{\Phi}&.
\end{split}
\end{equation}
The matrices $\mathbf{C}$ and $\mathbf{D}$ contain the coefficients $c$ and $d$ arranged in a way so as to represent the coupling sums (source terms) from \equ{cascade_equations_dicr} above. $\vec{\nabla}_\mu$ is the first derivative operator, $\vec{\mu}$ the mean energy loss or the stopping power of muons in dry air, 
with its energy dependence fully accounted for and arranged on the energy grid. The decay and interaction coefficients are obtained from Monte-Carlo simulations of particle interactions with air and of free decays. Interactions are simulated with \sibyll{-2.3c} and other cosmic ray interaction models, decays with \pythia{~8} \cite{pythia,Sjostrand:2015cx}. The computation is significantly accelerated by using sparse matrices and a method to reduce the stiffness \cite{cpc_paper}. Depending on the choice of models and the zenith angle, the solver needs between 0.1s and a few seconds on a single high-performance x86 core. \mceq{} is a relatively mature software that is gaining popularity among the neutrino communities and it has been used for several practical applications, \eg\cite{TheIceCube:2016oqi,Arguelles:2017eao,Edsjo:2017kjk}.

\subsection{Primary cosmic ray flux}
\label{sec:primary_flux_H3a}
For the discussion in this paper we choose one representative realistic model (called {\it H3a} \cite{Gaisser:2012em}) that contains the important features of the cosmic ray flux including the knee and the ankle. The origin of these features is understood in terms of transitions between classes of cosmic accelerators, such as supernova remnants, pulsars, gamma-ray bursts or active galactic nuclei. The flux of each mass component $\phi_i$ is modeled as a broken power-law, where each source class $j$ accelerates nuclei to a maximal cut-off rigidity $R_\text{c}$ 
\begin{equation}
  \phi_i(E) = \sum_{j = 1}^3 a_{i,j} E^{-\gamma_{i,j}} \times \exp\left [-\frac{E}{Z_i R_{\mathrm{c},j}}\right ].
\end{equation}

Our calculation method uses the superposition approximation, where the flux of interacting nuclei is expressed in terms of the individual nucleons (all-nucleon flux)
\begin{equation}
  \phi_\text{N} (E_\text{N}) = \sum_i A_i^2\ \phi_i(E_i= A_iE_\text{N})
\end{equation} 
or more precisely an energy dependent proton and neutron flux that can be obtained by substituting $A_i^2$ with $A_i Z_i$ or $A_i\text{N}$, respectively. 
\begin{figure}
  \includegraphics[width=\columnwidth]{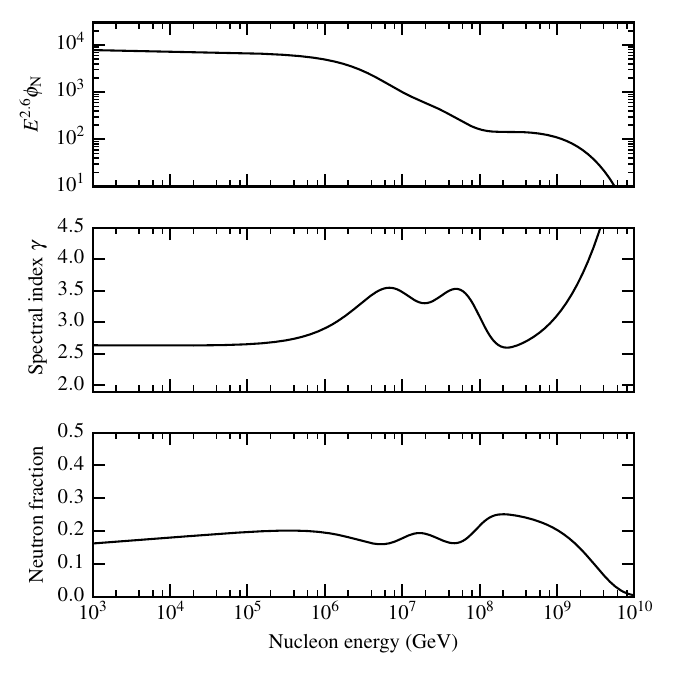}
  \caption{\label{fig:H3a} {\it Upper panel:} Flux of nucleons in $(\text{m}^{2}\ \text{s}\ \text{sr}\ \text{GeV}^{1.6})^{-1}$, converted from the flux of cosmic ray nuclei as predicted by the {\it H3a} model. {\it Middle panel:} spectral index, log-derivative of the upper curve. {\it Lower panel:} Fraction of neutrons in the nucleon flux. The spectral softening at around $1\,$PeV is an effect of the knee and the hardening at $\sim 100\,$PeV of the ankle of cosmic rays.}
\end{figure}
\figu{H3a} summarizes the different properties of this model.

\section{Connection between inclusive leptons and hadronic interactions}
\label{sec:had-prod}

A detailed study that relates the relevant properties of hadronic interactions to the spectra of atmospheric leptons is \cite{Sanuki:2006yd}. Here we revisit this topic using  the new \sibyll{-2.3c} event generator and the efficient numerical scheme, extending it to higher energies and to prompt leptons. All Figures and computations are made with \sibyll{-2.3c} if not otherwise noted.

\subsection{Distribution of cosmic ray energies}
\label{sec:prim-energies}
\begin{figure*}
  \includegraphics[width=0.85\textwidth]{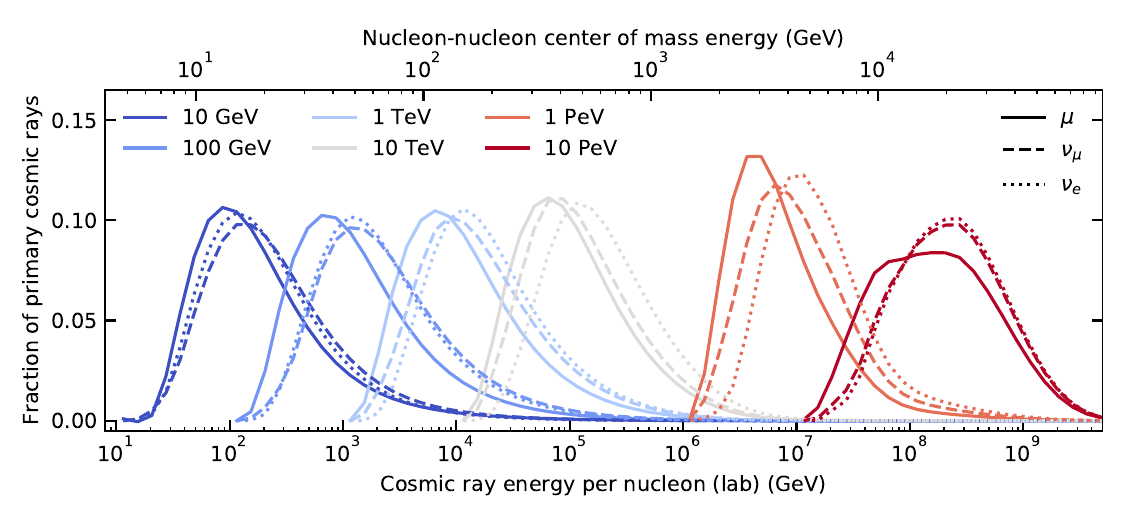}
  \caption{\label{fig:prim_e} Probability density functions (PDFs) of the primary nucleon energies corresponding to inclusive leptons a a fixed energy (colors). The solid, dashed and dotted lines refer to the individual lepton species. Atmospheric neutrinos up to $1\,$PeV probe center of mass collisions (top axis) that are within reach of current collider experiments (the LHC reaches $13\,$TeV).}
\end{figure*}
\begin{figure}
  \includegraphics[width=0.95\columnwidth]{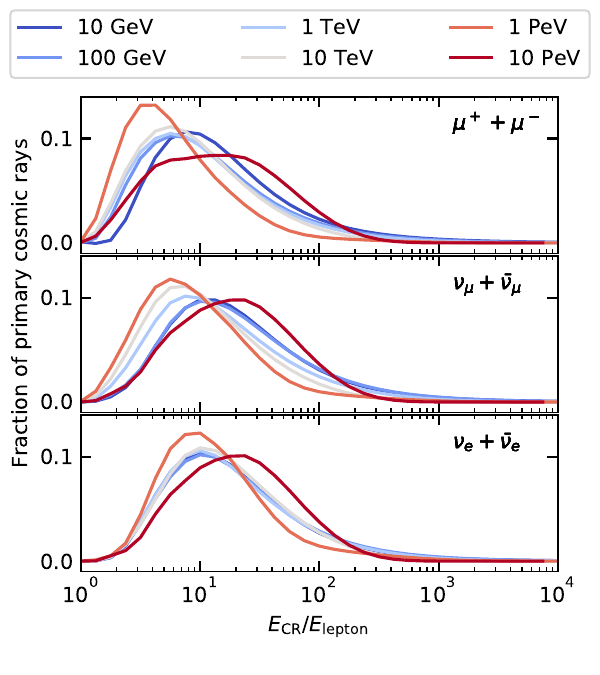}
    \caption{\label{fig:prim_e_fractional} Like \figu{prim_e} but as a function of fractional energy $E_{\rm cosmic\ ray}/E_{\rm lepton}$, for muons (upper panel), muon neutrinos (middle panel) and electron neutrinos (lower panel).}
\end{figure}
The energy spectrum of the primary nucleons that take part in the production of leptons at a fixed energy is shown in \figu{prim_e}. This probability density function (PDF) distribution can be obtained by calculating the contribution from each primary energy bin to the total inclusive flux. The distributions are replotted as a function of the energy fraction of the primary nucleon in \figu{prim_e_fractional}. The primary cosmic ray nucleon energies peak at $10\times E_\text{lepton}$ with a long tail extending to the highest energies, meaning that there is a non negligible probability that the primary cosmic ray can carry significantly more energy than an observed lepton. Since muons mostly originate from pion decays, they preserve a larger fraction of the momentum of the parent meson on average and therefore have a most probable primary energy that is somewhat lower than for neutrinos (around $8E_\mu$). This is a purely kinematical effect of the large muon/pion mass ratio. 

The shape of the distribution is significantly affected by the choice of the primary flux model, in particular by the position of the knee and the spectral indices of the all-nucleon flux. Further, it also depends on the type and the longitudinal spectrum of the mother meson. For this reason the shape of the primary energy distribution \revise{depends differently on energy for different leptons} as illustrated in \figu{prim_e_fractional}. 

At energies well below a PeV, the muons mostly originate from pions, and the shape is almost universal (blue lines in the upper panel of \figu{prim_e_fractional}), since $\rmd \sigma_{\pi^\pm}/\rmd x_\text{Lab}$ scales and the cosmic ray spectral index is constant. At higher energies, the spectral index of the primary flux becomes steeper since the cosmic ray spectrum is probed above the knee. Further, heavy flavor production becomes significant and the prompt flux dominates, explaining the large differences in the orange and red curves. For muon neutrinos (middle panel of \figu{prim_e_fractional}) the shape changes across the entire energy range. One of the reasons is that at lower energies the dominant mother particles are pions, at intermediate energies kaons and at the highest energies mostly D mesons. The different production and decay properties result in a larger variation for the peak cosmic ray energy between $5E_{\nu_\mu}$ and $20E_{\nu_\mu}$. We can conclude that for conventional muons the relation to the primary nucleon energy scales, but not if prompt fluxes are taken into account. For muon neutrinos the scaling assumption is not accurate. Electron neutrinos originate mostly from charged and neutral kaon decays, which have similar production and kinematics. The distributions look similar to the muon case, peaking at $10E_{\nu_\mu}$ at intermediate energies. 

It is interesting to note that the corresponding center-of-mass (\cm{}) energies, relevant for atmospheric lepton production up to multi-PeV energies, are within reach of current particle colliders. However, it is very challenging to create and operate a detector technology that is close enough to the beam to cover the relevant (as we discuss later) longitudinal momentum range with $\xl$ or $x_{\rm F} > 0.2$ \footnote{Feynman-$x$ $x_{\rm F}=p_{z}^{\rm c.m.}/p_{z,\text{max}}^{\rm c.m.}\simeq 2p_{z}/\sqrt{s}$}. Another obstacle is that nuclear interactions at TeV energies are currently probed in lead-lead or proton-lead collisions at the LHC, but lighter ions in the mass range of air molecules are accessible only in simulations.

\subsection{Relevant hadrons for inclusive lepton production}

\label{ssec:relevant_hadrons}
\begin{figure*}
  \includegraphics[width=0.7\textwidth]{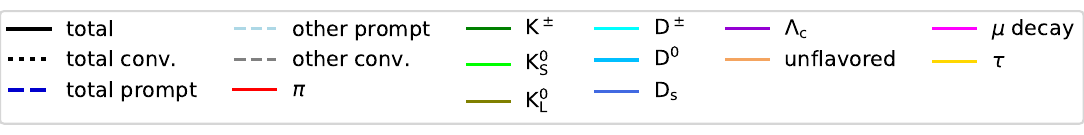}
  \includegraphics[width=0.9\textwidth]{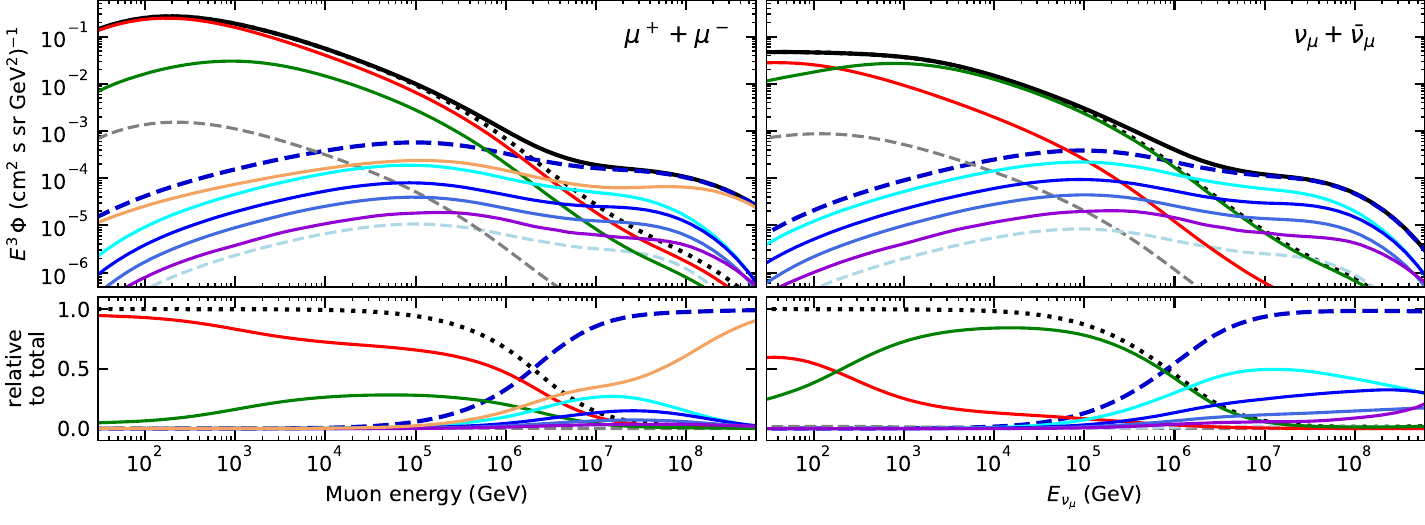}
  \includegraphics[width=0.9\textwidth]{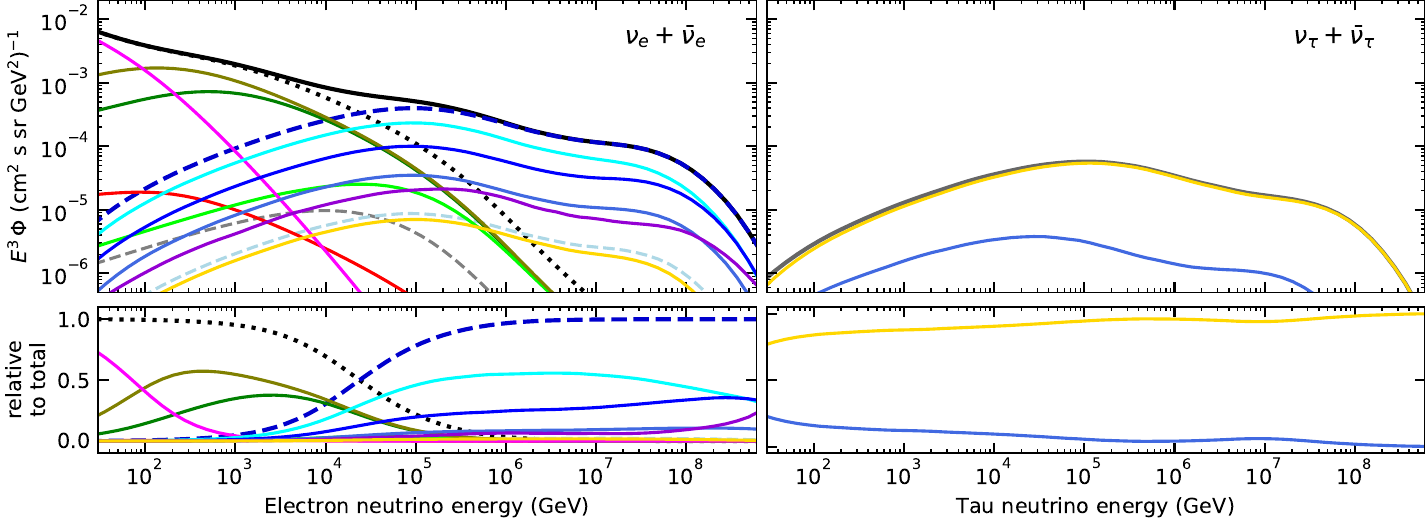}
  \caption{\label{fig:hadron_contribution} Contribution from decays of various particles to the atmospheric $\mu^+ + \mu^-$ (top left), $\nu_\mu + \bar{\nu}_\mu$ (top right), $\nu_e + \bar{\nu}_e$ (bottom left) and $\nu_\tau + \bar{\nu}_\tau$ (bottom right) flux in \sibyll{-2.3c} and {\it H3a} primary model at $\theta =60^\circ$.}
\end{figure*}
We continue discussing the connection between hadronic interactions and inclusive leptons by looking at the different hadron species that give rise to a sizable production of inclusive leptons. Most atmospheric leptons originate from weak and partly from electromagnetic decays of the most abundant mesons, \ie{} charged pions and kaons. Two aspects of particle production are relevant here; the longitudinal production spectrum, such as the $p_T$ integrated differential cross section, and, the energy distribution among the decay products and their inclusive branching ratios.

The hadronization routines in \sibyll{} can essentially produce all relevant hadrons and resonances up to masses of the $\Omega_\text{ccc}$ baryon. Inclusive $p_T$-integrated cross sections $\rmd \sigma_h/\rmd x_\text{Lab}$ are computed for each hadronic species irrespective of its life time. Decays are tabulated separately by using \pythia{} 8 \cite{pythia,Sjostrand:2015cx}. The inelastic interaction cross sections of more exotic hadrons are assumed to be equal to $\sigma^\text{prod}_\text{nucleon-Air}$ for all baryons, $\sigma^\text{prod}_{\pi^\pm - \text{Air}}$ for pions and light mesons, and $\sigma^\text{prod}_{\text{K}^\pm \text{-Air}}$ for heavier mesons including charmed. 

The various groups of mother particles that directly decay into leptons and contribute to the inclusive flux are shown in \figu{hadron_contribution}. Sub-leading contributions are summed together in the ``other'' groups. As the energy increases, the decays of particles become suppressed above the critical energy. Heavier and less abundant hadrons dominate at very high energy and produce prompt atmospheric leptons. 
\begin{figure}
  \includegraphics[width=\columnwidth]{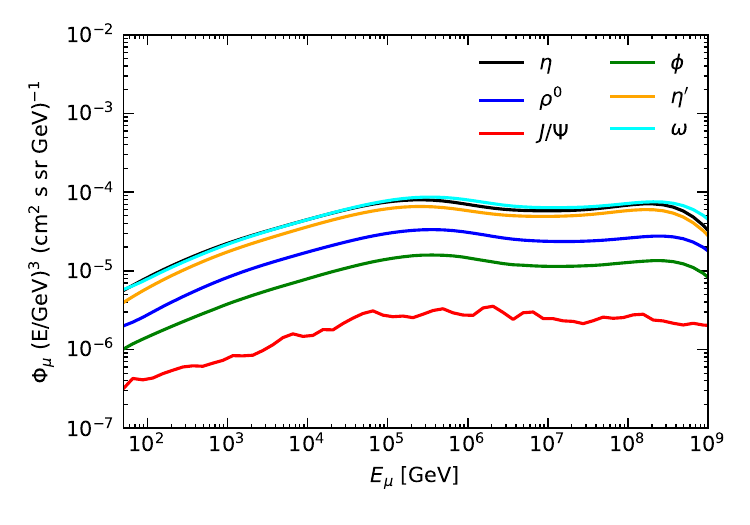}
  \caption{\label{fig:unflavored} Break-down of the unflavored component of the prompt muon flux.}
\end{figure}
As expected, the conventional muons stem from the decays of charged pions and, with a smaller contribution, from charged kaons (upper left panel in \figu{hadron_contribution}). Prompt muons have two sources, decays of charmed mesons and a component from electromagnetic decays of unflavored mesons \cite{illana_eta_2010}. The detailed break-down of the contributors to the unflavored component is shown \figu{unflavored}. A contribution at a similar level as charm comes from the process $\eta~\text{and}~\eta' \to \mu^+ \mu^- \gamma$, breaking the correlation between prompt fluxes of muons and neutrinos. The cross-over between conventional and prompt flux happens at several PeV and depends on the choice of models and the zenith angle. Further sources of high energy muons that are not included in our calculation are the photo-production of muon pairs, which is suppressed by $10^{-4}$ wrt.\ the pair production cross section $\sigma_{e^+ e^-}$\cite{Nelson:1985ec}, and the nuclear interactions of muons. While the muon pair production can significantly contribute to inclusive fluxes at very high (PeV) energies, the nuclear interactions are only important for the low energy tail of muon bundles in air showers.

At E $\gtrsim100\,$GeV the main source of muon neutrinos (upper right panel) are semi-leptonic and 3-body decays of charged kaons, see \eg{}\ \cite{Honda_1D_1995} for a more detailed discussion of relevant channels. Pion and muon decays dominate below this energy. Prompt neutrinos originate from decays of charged and neutral D-mesons, where the fluxes from D$^\pm$ are a factor of three higher. Since pions very rarely decay into electron neutrinos (lower left panel), those come mostly from decays of neutral and charged kaons. At energies below $100\,$GeV and for near-horizontal zenith angles the dominant fraction of electron neutrinos is from muon decays, resulting in a strong association with the muon flux. In turn, this means that the precision of the electron neutrino prediction for a few to several tens of GeV is linked to the modeling of pion production and muon energy loss and, to a lesser extent, to kaon production.

Atmospheric tau neutrinos (lower right panel) are rare \cite{Bulmahn:2010pg}, but we can discuss their flux for completeness. The dominant production channel of tau neutrinos is the decay of $\text{D}_\text{s}^+ \to \tau^+ + \nu_\tau$, where the subsequent decay of $\tau \to \nu_\tau + X$ is more efficient in producing a forward tau neutrino, than the decay of the meson. Therefore most of the tau neutrino flux comes from the decay of the tau lepton itself (black and blue line in lower right panel in \figu{hadron_contribution}).

Other sources of atmospheric leptons that are not taken into account in our calculation are B-hadrons. Their contribution to the prompt flux can be of the order of 10\% \cite{2003AcPPB..34.3273M,2009JCAP...09..008I}.

\subsection{Muon charge ratio}
\label{ssec:charge_ratio}
\begin{figure}
  \includegraphics[width=0.9\columnwidth]{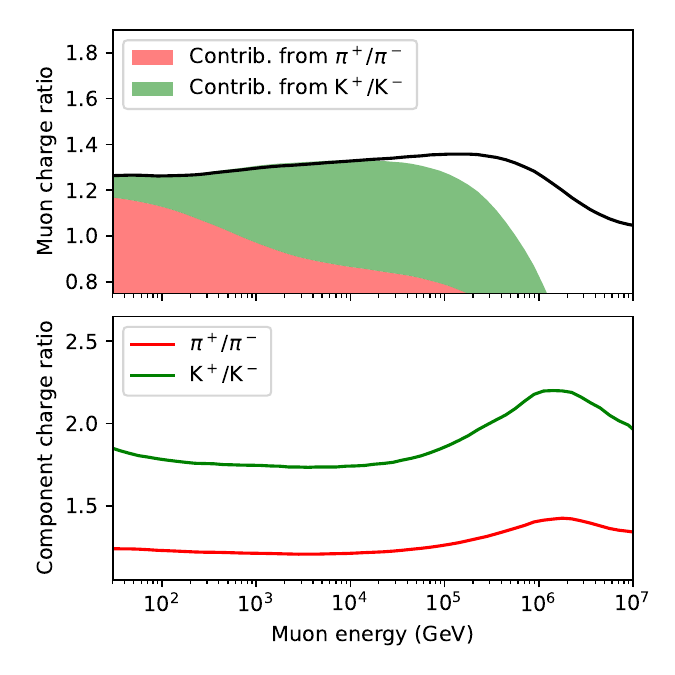}
  \caption{\label{fig:charge_ratio_simple} Muons charge ratio and effective components from pion and kaon decays. (Upper panel) The solid line is the total muon charge ratio from \sibyll{} and the areas the individual weights from the pion and kaon component. (Lower panel) The colored areas are a measure of the effective weight of the pion and the kaon component. The variations at $1\,$PeV come from the change of the spectral index at the knee, resulting in a different weight (or mean) of the $\xl$ distribution of secondaries. }
\end{figure}
The inclusive muon charge ratio $R_\mu$ has been perceived in the literature as \revise{an observable with a high} sensitivity to the hadronic physics in atmospheric cascades \cite{Gaisser:2012em,Adamson:2007ww,Agafonova:2014mzx}. $R_\mu$ is approximately 1.25 below $1\,$TeV and increases to slightly above 1.35 at higher energies. The reason for this transition can be seen in \figu{charge_ratio_simple} as a transition from an energy range, dominated by the charge ratio of the pion component, to a higher energy range where kaons become more important.

Direct constraints on the production of pions or kaons in proton-air interactions from $R_\mu$ directly are difficult to derive, since the spectral index and the neutron fraction in the cosmic nucleon flux introduce some degeneracies. While the charge ratio $R_\mu(E_\mu)$ at a fixed $E_\mu$ covers a range of primary interaction energies (compare with \figu{prim_e}), the forward particle spectra approximately scale (see Section \ref{ssec:sib23_leading} and \figu{sib-scaling}), alleviating this problem in the interpretation in terms of hadronic interactions. More important is the degeneracy between the shape of the particle production spectrum and the spectral index of primary cosmic rays, or even the spectrum of secondary projectiles downstream of the air shower. As we discuss below, a sizable fraction of inclusive muons stems from these secondary interactions and thus changes as a function of depth. 

\equ{z-factor} demonstrates that the size of the pion and the kaon component is controlled by the convolution of the projectile spectrum and the particle production spectrum (see \equ{z-factor}). Since the charge ratio is not constant along $\xl$ (see \figu{sib-na49-pion-charge-ratio}), $R_\mu$ can only constrain the weighted integral \equ{z-factor} and not the the spectrum of leading mesons directly. This implies that in order to access the microscopic hadronic physics through inclusive lepton measurements one necessarily needs to take multiple correlated observables into account, like the angular distributions of muons and $R_\mu$ together with the inclusive muon neutrino and electron neutrino fluxes.

\subsection{Inclusive muons vs.\ muon bundles}
\begin{figure}
  \includegraphics[width=0.9\columnwidth]{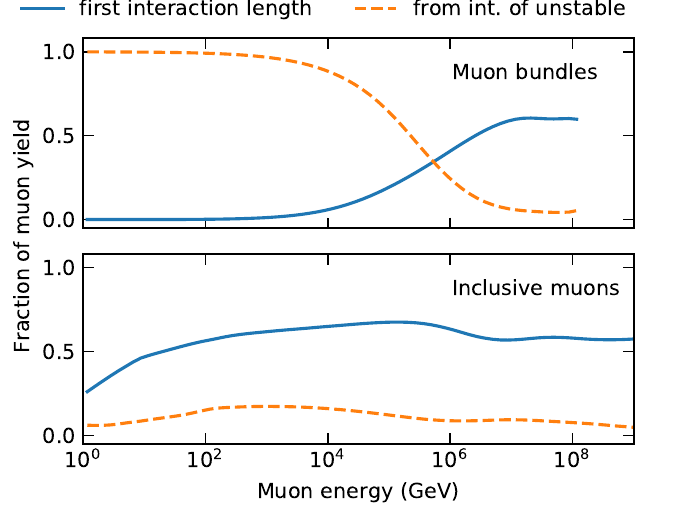}
  \caption{\label{fig:sec_interactions} Red dashed lines show the contribution to the muon spectrum by interactions of unstable particles ($\pi^\pm$, $\Lambda$, \etc) and solid blue the fraction of muons that is related to hadronic interactions in the first interaction length ($\sim 90\,$g$/$cm$^2$) of the cosmic ray nucleon. The upper panel is computed for a single $100\,$PeV proton and a vertical incidence angle.}
\end{figure}
\revise{The term muon bundles refers to the muon signature that can be observed by underground detectors, such as L3+c, MINOS or IceCube. High-energy air-showers produce large numbers of muons at ground, distributed from below $1\,$GeV up to an energy comparable with the primary cosmic ray. Most energy is concentrated in a small cone around the shower core, which is aligned with the cosmic ray direction. The overburden, rock for underground detectors or water/ice for Cherenkov neutrino telescopes, absorbs the low energy muons and a small fraction between one and several hundreds of muons reaches the detector.

Inclusive muon fluxes are (``experimentally'') obtained by scoring the momentum of each individual muon and integrating over time. The time integration translates into an integration over the cosmic ray energy spectrum at the top of the atmosphere. The muons in a muon bundle come from the same air-shower. 

\figu{sec_interactions} demonstrates that bundles and inclusive fluxes depend differently on hadronic interactions. The contribution from (secondary) interactions of unstable particles, such as pions, kaons or $\Lambda$ baryons is very high for bundles or muons in air-showers. Whereas for the inclusive fluxes the integration over the steep primary cosmic-ray spectrum results in a suppression of the importance of secondary interactions. \figu{sec_interactions} also shows that observations of muon bundles become sensitive to the properties of the first few high-energy interactions if the energy cutoff is high enough. While detectors, such as the Pierre Auger and the IceCube Observatories, \revise{both look} at bundles, the relevant phase space for hadronic interactions is different. \revise{Muons observed at the surface by Auger or IceTop are} dominated by muons from low-energy $\pi$-air interactions.  \revise{On the other hand, $\sim$TeV muons observed in the deep array of IceCube are} more sensitive to high-energy primary interactions.}

\subsection{Hadrons that do not decay into leptons}
\begin{figure*}
  \includegraphics[width=0.95\textwidth]{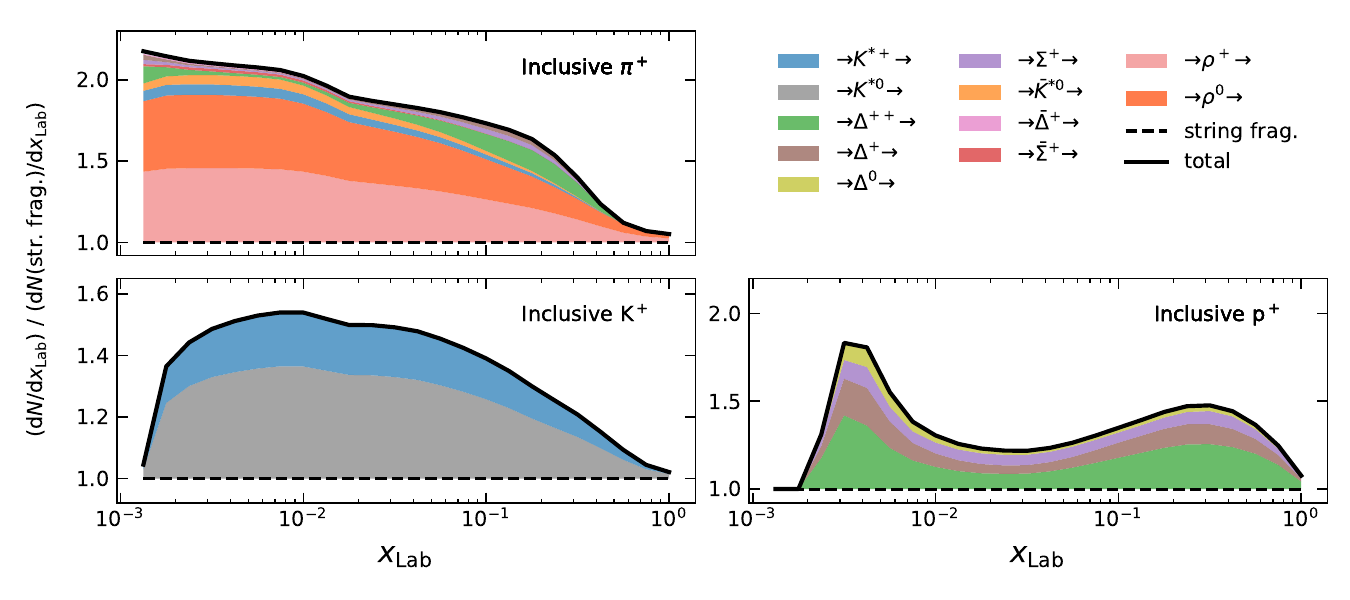}
  \caption{\label{fig:feed_down} Stacked plot of the feed-down contributions of short-lived hadrons to the inclusive pion, kaon and proton spectra, computed with \sibyll{} at $E_{\rm lab} = 1\,$TeV for proton-Carbon interactions. The hadrons produced directly in string fragmentation or resonance decay are normalized to one (dashed line). Hadrons that contribute below a few percent are omitted.}
\end{figure*}
In the previous sections, we discussed the role of hadrons with sizable leptonic branching ratios that can directly contribute to the inclusive flux. Here, we outline why an accurate description of the other particles (with rare leptonic decays) is indispensable in inclusive flux calculations, how these hadrons are related to the conventional pion and kaon components.

The unstable \revise{hadrons} in atmospheric cascades decay into the lightest unstable species ($\pi^{\pm}$, K, etc.), which ultimately decay into leptons or nucleons before reaching the ground. The life time of particles (for instance) $\rho$-mesons is so short that their  explicit presence in the transport equations has little impact on the development. Through the decay into two pions ($\text{BR} \sim 100\%$) the $\rho$ mesons feed down into the secondary spectrum of pions. Therefore, they have to be taken into account either explicitly (in the transport equations) or implicitly in the inclusive production spectrum of pions. As it has been discussed in \cite{Drescher:2007hc,Ostapchenko:2013pia}, leading $\rho$ meson in pion-nucleus interactions (instead of leading $\pi^0$ that feed the electromagnetic cascade) can significantly affect the development of the muon content in an extensive air shower.

\figu{feed_down} shows the feed down from vector mesons, strange baryons and resonances to the inclusive yields of pions, kaons and protons in p-Air interactions. These additional channels are remarkably large, demonstrating the importance of choosing accurately the definition of stable/resolved particles when comparing and tuning event generators to accelerator data or to other models. This example clearly demonstrates why the development of interaction models requires significant effort and why there are no accessible ``knobs'' for ``tuning'' a model to, for example, pion measurements. An enhanced production of light mesons would necessarily lead to reduced production of vector mesons or baryons, creating tension with other observables. As described in Section \ref{sec:sib23}, \sibyll{} uses the {\it Lund} string fragmentation model \cite{Bengtsson:1987kr} that consistently connects the production of light and heavier mesons to the available energy in color strings using a set of phenomenological probability parameters. Simply speaking, in an ideal world these parameters should be universal and can be derived from measurements at lepton colliders. However in practice this is not always true, in particular in hadronic collisions at very high energies where additional non-perturbative methods are invoked to describe hadronization \cite{Sjostrand:2015cx}, at least when sticking to the fragmentation of color strings as the underlying picture for hadron formation.

\subsection{Angular distribution}
\label{ssec:angular_dist}
\begin{figure*}
  \includegraphics[width=\textwidth]{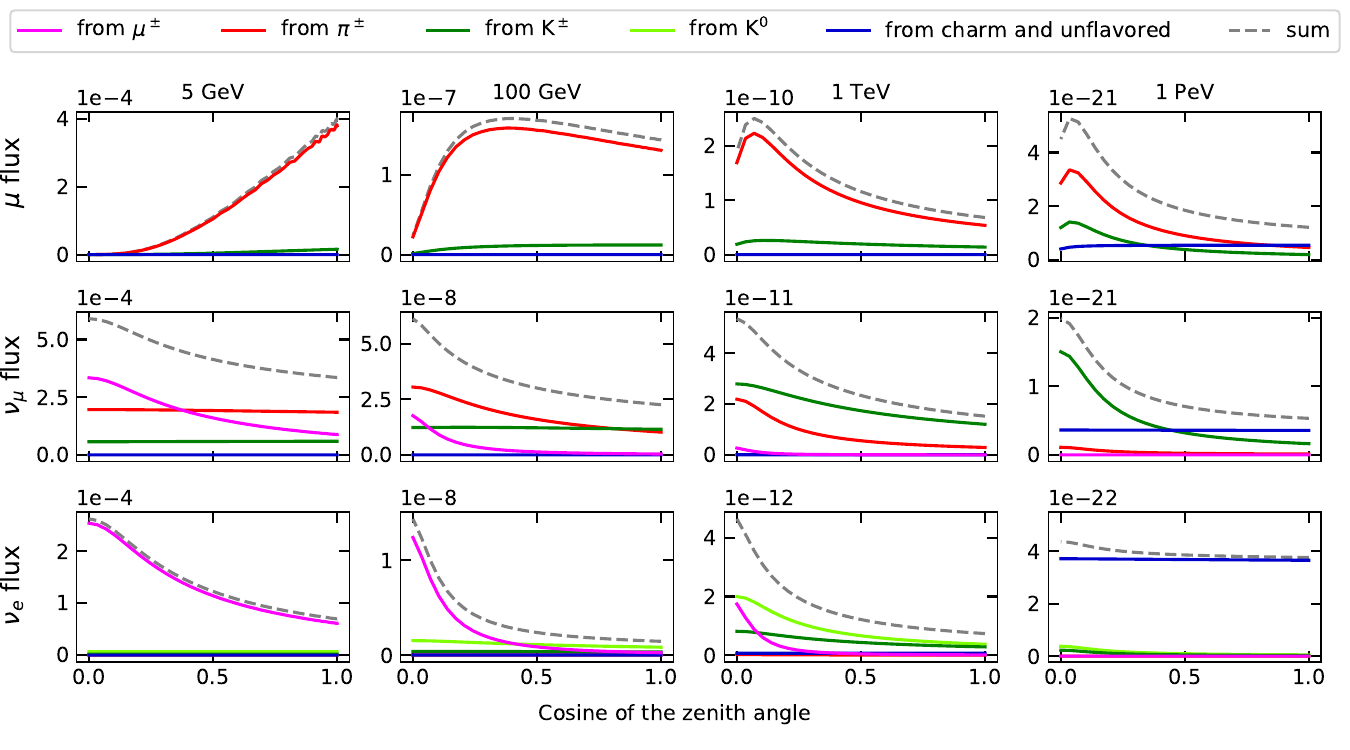}
  \caption{\label{fig:angular_distribution} Zenith angle dependence of the hadronic components of the atmospheric flux in (cm$^2$ s sr GeV)$^{-1}$. The signature of a ``prompt'' component is a flat zenith distribution, meaning that these ``mother-mesons'' do not interact with the atmosphere before decay. At energies above the critical energy $\epsilon$ this probability increases and depends on the density gradient along the cascade trajectory. The gradient is smaller for horizontal trajectories, thus the probability for unstable particles to decay becomes higher again. The prompt component from charmed and unflavored meson decay is flat up to the highest energies due to the short life times these hadrons.}
\end{figure*}
The panels of \figu{angular_distribution} show the dependence of the differential fluxes of leptons on the zenith angle, as it would be perceived by a detector located at the surface. Traditionally, this behavior is described by a combination of the steepening of a spectral components (from $\mu$, $\pi$ or K decay) that starts above different critical energies (see \equ{semi-anal}). The critical energies are approximately 1, 115 and $850\,$GeV for $\mu$, $\pi$ and K mothers, respectively \cite{Gaisser:2016uoy}. These energies represent the transition from a decay to an interaction dominated transport and in reality they depend on the atmospheric model, the production height and the zenith angle. Muon decay is an additional process that affects the angular distribution of muon and neutrino fluxes at low energies, where a deviation from the typical $1/\cos{(\theta)}$ shape can be seen in the upper left panel, and even at higher energies for very extreme angles.

The strong enhancement of near-horizontal fluxes originates from the suppression of secondary interactions of mesons, since the bulk of high energy particles from the first interactions are traversing a very thin atmosphere and do not accumulate enough depth. The distinctive feature of the prompt component is the flat angular distribution. For fluxes from charmed meson decays, re-interactions with the atmosphere become relevant above $5$-$10\,$PeV (not shown in \figu{angular_distribution}) leading to deviations from the flat distributions.

At very high energies the flux is not always dominated by prompt leptons. As the right panels clearly show, the prompt component is dominant for vertical directions but becomes sub-dominant towards the horizon. For the detection of prompt fluxes free from the contamination with the conventional component, it would be necessary to search in electron neutrinos at several hundreds of TeV, or, in muon neutrinos above PeV energies, preferably in vertical directions. With present neutrino telescopes, such as IceCube, ANTARES and GVD, the detection is extremely challenging since the prompt flux is low, and, diluted by backgrounds from muon bundles in the potentially interesting channels (vertical down-going or electron neutrino cascades).

\subsection{Particle production phase space}
\label{ssec:phasespace}
\begin{figure}
  \includegraphics[width=0.9\columnwidth]{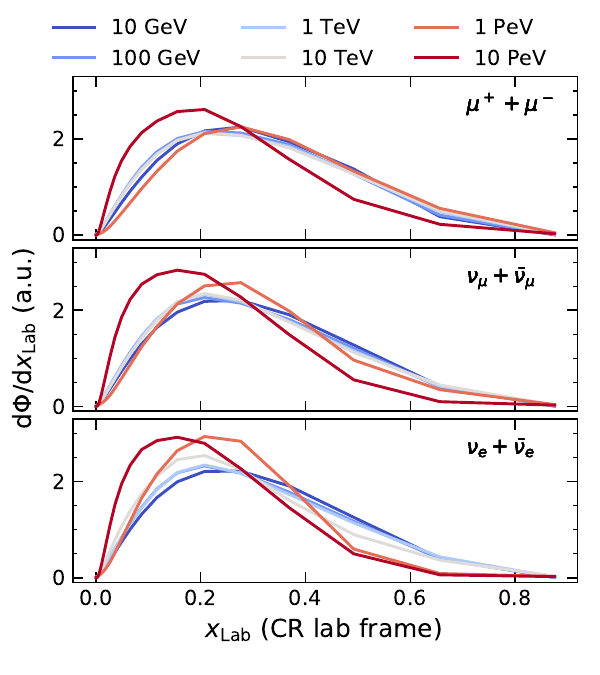}
  \caption{\label{fig:xl-phasespace} PDF of the relevant $\xl$ values for inclusive leptons at fixed energy. These curves are derived from the inclusive flux, \ie{} integrated over the cosmic ray spectrum. An energy independent Z-factor approach would results in a universal shape. Here, the energy dependence comes from deviations from perfect Feynman scaling and from changes in the spectral index of the primary nucleon flux due to the knee and ankle.}
\end{figure}
In section \ref{sec:prim-energies}, we outlined that atmospheric leptons probe hadronic interactions at center-of-mass energies that are accessible to recent accelerators and we showed that the most probable energy of the incident cosmic ray is $E_\text{CR} \sim 10E_\mu$ or very approximately at 
\begin{equation}
  \xl \sim \frac{E_\pi}{E_\text{CR}} = \frac{E_\mu / 0.6}{E_\text{CR}} \approx 0.16,
\end{equation}
where the 0.6 is the average energy fraction transferred to muons in pion decays. The cascade evolution implies additional factors such as secondary interactions, contributions from other decay channels the decrease of the nucleon energy in subsequent interactions etc. A PDF for $\xl$ values that contribute to the flux of leptons at a fixed energy is shown in \figu{xl-phasespace}. The energy dependence of these curves can be explained analogously to the energy behavior of \figu{prim_e_fractional} and comes from deviations from ideal Feynman scaling of the interaction model and the knee and ankle in the cosmic rays. Note that the $\xl$ values commonly refer to all hadrons including (p, n, $\pi$, K). 

A value of $\xl = 0.3$ for an incident proton at 10 PeV, for example, implies that the scattering angle of the secondary particle is of the order of few $\mu$rad and impossible to detect at a particle collider experiment without dedicated detectors. At high energies, the hadronic interactions are only accessible through simulation, and thus constitute the dominant source of uncertainty in the calculations of inclusive fluxes.

\section{Hadron production in \sibyll{~2.3c}}
\label{sec:sib23}
The hadron interaction model \sibyll{} is designed mainly for the use
in cosmic ray air shower simulations. While the general features of
QCD like quark confinement, multiple interactions and jet production
are included in the model, particular features that are relevant for
the development of air showers, like diffraction dissociation and
forward particle flow are implemented in more detail.

The interaction model in \sibyll{} is based on the two-component dual
parton model with soft and hard minijets~\cite{Capella:1992yb}. It also includes
low- and high mass diffraction and a model for the excitation of beam
remnants~\cite{Riehn:2015oba,Ahn:2009wx}. The hadronization model is
based on the Lund string fragmentation model~\cite{Bengtsson:1987kr,
  Sjostrand:1987xj}. Hard scattering is distinguished from soft
scattering by a cutoff in transverse momentum. The cross section for
hard scattering is calculated to LO in QCD at the scale defined by the
cutoff in $p_{\rm T}$. Saturation is included by means of an energy
dependent increase of the $p_{\rm T}$ scale. Contributions from quarks
of all flavors and gluons are included with their full kinematics as
determined by the parton distribution functions. However, in the
subsequent fragmentation of the scattered partons, no distinction
between the different flavors is made. The string (color-flow)
configuration of all the parton interactions are treated as gluon
gluon scattering (see \figu{sib-minijet-charm} for
illustration). The soft interaction cross section is modeled with a
parameterization based on the Regge field theory~\cite{Donnachie:1992ny}.

Two aspects of hadron interactions are improved in the latest
version of \sibyll{} motivated by the discussion above. These are
the production of leading particles and the production of charmed
hadrons. The new treatment of leading particles of the remnant
model is discussed in Sect.~\ref{ssec:sib23_leading})
and the charm model is covered in Sect.~\ref{ssec:sib23_charm}.

\subsection{Leading particles}
\label{ssec:sib23_leading}

\begin{figure*}
  \includegraphics[width=\columnwidth]{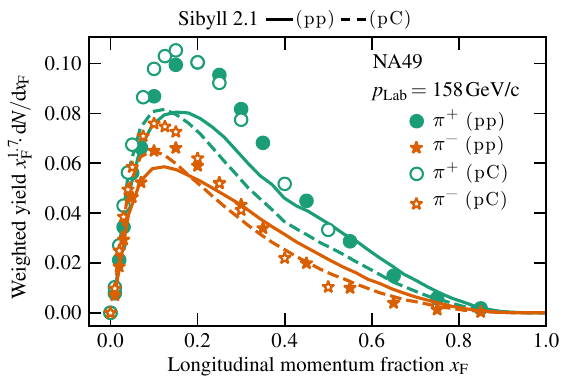}
  \hfill
  \includegraphics[width=\columnwidth]{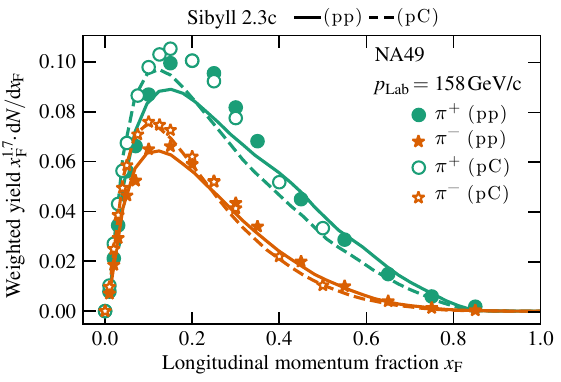}
  \caption{\label{fig:sib-na49-pions}NA49 pions in \sibyll{-2.1} (left) and \sibyll{-2.3c} (right)~\cite{Alt:2005zq,Alt:2006fr} for proton-proton (solid) and proton-Carbon (dashed) interactions. In the fragmentation region $\pi^+$ (u$\bar{\mathrm{d}}$) dominate over $\pi^-$ (d$\bar{\mathrm{u}}$) because of the quark content of the proton (uud). The remnant model in \sibyll{-2.3c} is tuned to reproduce these data in the forward region.}
\end{figure*}

\begin{figure}
  \includegraphics[width=\columnwidth]{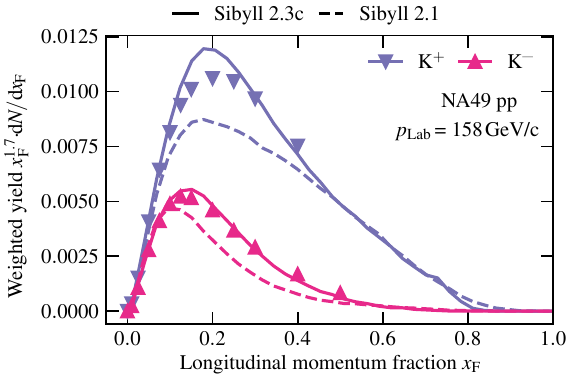}
  \caption{\label{fig:sib-na49-kaons}Longitudinal momentum spectrum of charged kaons measured in pp fixed-target collisions by NA49~\cite{Anticic:2010yg}, compared with \sibyll{}.}
\end{figure}

\begin{figure}
  \includegraphics[width=\columnwidth]{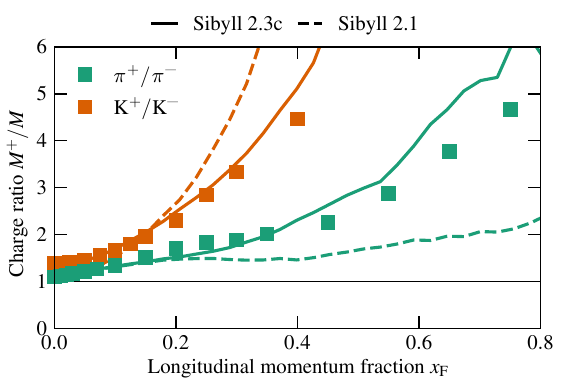}
  \caption{\label{fig:sib-na49-pion-charge-ratio} NA49 measurement of the charge ratio of pions and kaons in pp interactions as a function of Feynman-$x$~\cite{Alt:2005zq,Alt:2006fr,Anticic:2010yg}, compared with \sibyll{}.}
\end{figure}



\begin{figure*}
  \includegraphics[width=\columnwidth]{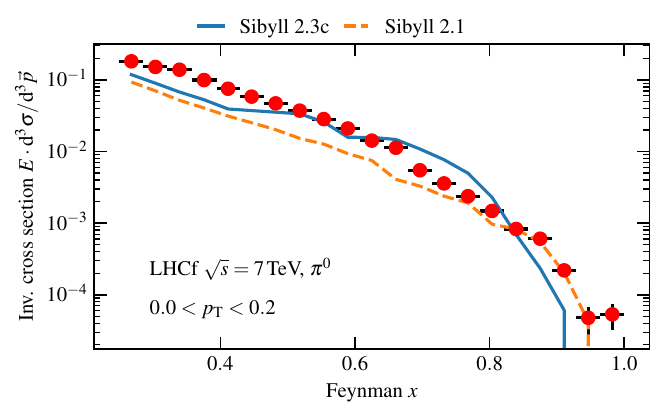}
  \hfill
  \includegraphics[width=\columnwidth]{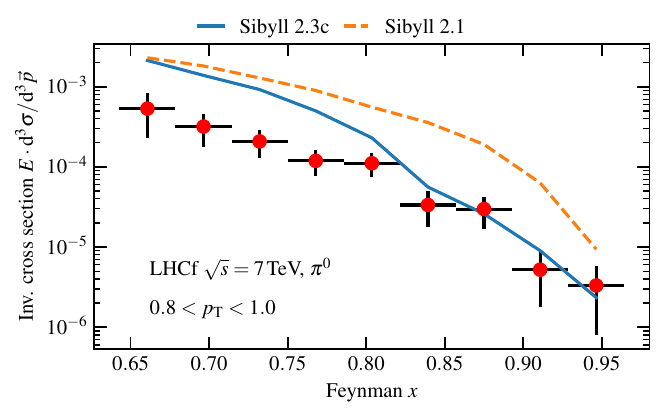}
  \caption{\label{fig:sib-lhcf-pi0}Longitudinal momentum spectrum of the  of neutral pions at $7\,$TeV. Data are from the LHCf spectrometer, covering $8.9<\eta<10.2$ in two different $p_T$ bins.}
\end{figure*}

\begin{figure}
  \includegraphics[width=\columnwidth]{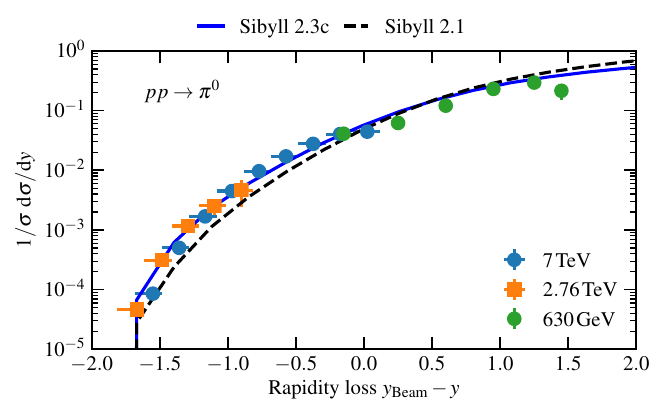}
  \caption{\label{fig:pi0-scaling}Yield of neutral pions as a function of the rapidity loss $y_{\rm loss}=y_{\rm beam}-y$ at different \cm{} energies~\cite{Adriani:2015iwv,Pare:1989mr}. If Feynman scaling holds in the fragmentation region, the spectrum is universal.}
\end{figure}

\begin{figure*}
  \includegraphics[width=\columnwidth]{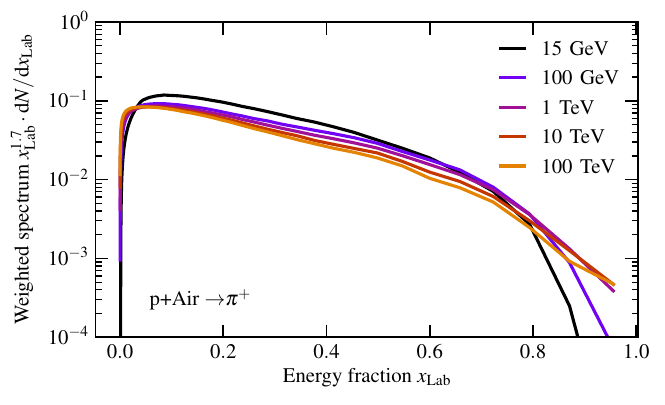}
  \hfill
  \includegraphics[width=\columnwidth]{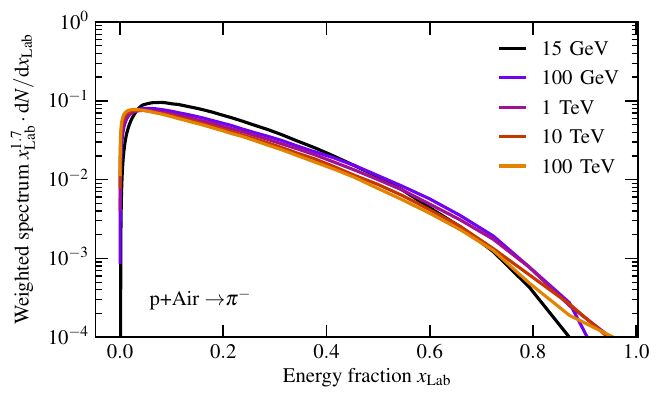}
  \vfill
  \includegraphics[width=\columnwidth]{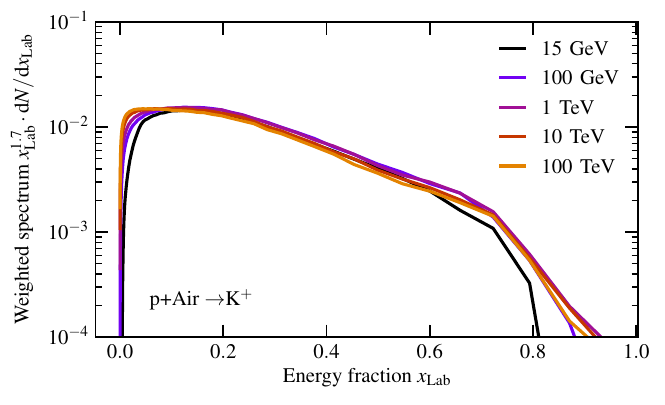}
  \hfill
  \includegraphics[width=\columnwidth]{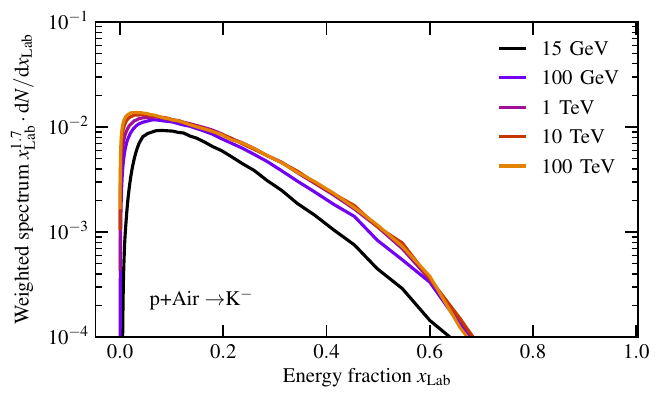}
  \caption{\label{fig:sib-scaling} Scaling behavior of the energy distribution of secondary mesons in \sibyll{-2.3c}. For perfect scaling the lines for the different laboratory frame projectile energies should be on top of each other. The lowest energy line deviates due to threshold. At very small $\xl$ scaling violations due to multi-parton interactions are expected and visible at higher energies. For $\pi^-$ and K$^-$ the scaling is better than for their positively charged counterparts, which are more affected by the remnant model.}
\end{figure*}

While the particle production in forward phase space ($\xl > 0.2$) plays an important role for the fluxes of leptons in the atmosphere (see Sect.~\ref{ssec:phasespace}), it is even more important for the charge and flavor ratios (see Sect.~\ref{ssec:charge_ratio}). The reason is that the weight applied on the longitudinal spectrum emphasizes the forward phase space (\equ{z-factor}), where the flavor asymmetry is strongest. This forward asymmetry is also known as leading particle effect and it is related to the valence quarks dominating the momentum distributions of hadrons in soft interactions.

In \sibyll{} the valence quarks are assumed to undergo a soft interaction (two-string model), where each hadron is split into two valence partons (baryon: quark and di-quark, meson: quark and antiquark) and two color strings are formed between the partons of the two hadrons. The basic version of this non-perturbative configuration corresponds to
the upper baryon in \figu{sib-leading-charm}.  The nature of the leading
baryon is determined by the flavor of the quark (from the $q\bar{q}$ pair) with which
it recombines.  (The lower baryon in \figu{sib-leading-charm} illustrates
remnant splitting channel to be discussed in the next paragraph.)  In the standard
fragmentation of a projectile baryon,
the fraction of the momentum carried by the valence quark is assumed to follow
\begin{equation}
  f_{\rm q}(x) = (1-x)^{3} \, (x^2+0.3 \mathrm{GeV}^2/s \, )^{-1/4} \label{eq:soft-x}
\end{equation}
and the di-quark (quark or antiquark for mesons) is assigned the remaining momentum $x_{\rm diq}=1-x_{\rm q}$. In \sibyll{-2.1} leading particles emerge from the strings due to the large momentum fractions of the di-quark in combination with a hard fragmentation function used only for the hadrons produced at the string ends that are related to the beam particle. 

In \sibyll{-2.3c} the hard fragmentation is replaced by a beam remnant model~\cite{Drescher:2007hc,Liu:2003we,Sjostrand:2004pf}, where the valence quarks are separated from the sea partons that fragment as an independent system. The configuration of strings in the presence of a remnant is illustrated in the lower baryon line of \figu{sib-leading-charm}. The fraction of momentum assigned to the remnant is taken from $f_{\rm r}(x)=x^{1.5}$ and the excitation mass is distributed $\propto M_{\rm r}^{-3}$ where the lower limit is the hadron mass and the maximal mass is $M_{\rm r}^2 \, s^{-1}=0.02$. The energy required for the excitation is transferred from the other hadron.  

To account for the possible absorption of the valence quarks in the scattering process the remnant is formed with the constant probability of 60\%, suppressed by the number of soft and hard parton interactions and, in case of nuclear interactions, with the number of nucleons involved
\begin{equation}
  P_{\rm r} = P_{\rm r,0}^{\,N_{\rm w}+0.2 \, (n_{\rm soft}+n_{\rm hard})} \ , \nonumber
\end{equation}
with $P_{\rm r,0}=0.6$. Through this mechanism the leading proton distribution obtains the characteristic dip in the transition from hadron to nuclear targets, while the meson spectra are not affected. At the same time the charge ratio of leading pions can be described more accurately as the fragmentation of the proton remnant through nucleon resonances. The small strings preserve the isospin of the proton and favor the production of $\pi^+$ (see \figu{sib-na49-pions}). For kaons the charge ratio is affected even more strongly (\figu{sib-na49-kaons}) as the nucleon can only transition into a hyperon and a positive charged kaon, \eg{}
\begin{equation}
N^\star(\mathrm{uud}) \to \mathrm{u}\bar{\mathrm{s}} + \mathrm{sud} \to \Lambda^0 + \,
\mathrm{K}^+ \  . \nonumber
\end{equation}
This model yields a viable explanation for the effect of associated production, in which both $\Lambda$ and K$^+$ exhibit a leading particle effect. The models, in which the strings span between the valence quarks and $s\bar{s}$ pairs from the sea do not explain such a strong forward enhancement, since kinematically the strange hadrons form centrally. In \figu{sib-na49-pion-charge-ratio} the resulting charge ratios for pions and kaons are compared with measurements in NA49~\cite{Alt:2005zq,Alt:2006fr,Anticic:2010yg}. The new model describes the pion charge ratio well. For kaons the description has improved but towards large $x_{\rm F}$ the ratio is still overestimated.

Unfortunately there are no data on leading meson production and charge ratios available at $\sqrt{s} > 50\,$GeV to directly test the model. Indirectly the model is constrained by the spectrum of leading neutral pions measured at the very forward spectrometer LHCf~\cite{Adriani:2015iwv}. While the model reproduces the spectral shape in $x_{\rm F}$, the transverse momentum distribution is not described as well (\figu{sib-lhcf-pi0}). 

LHCb~\cite{Aaij:2012ut}, the detector that provides particle identification with the
largest rapidity coverage at the LHC, only covers a small region of longitudinal phase space $x_{\rm F} \ll 0.1$ and therefore can not constrain the leading charge
ratios. In a proposed fixed target configuration for LHCb~\cite{Lansberg:2012wj,Ulrich:2015uwb}, the charge ratios in the forward region could be determined for $\sqrt{s}=104\,$GeV. By comparing with \figu{prim_e_fractional}, this would correspond to the charge and flavor ratios of muons and neutrinos in the TeV to PeV range.

One of the central assumptions in analytical calculations of the atmospheric fluxes is the scaling behavior (energy independence) of the particle production spectra at large Feynman-$x$ with energy, the so-called Feynman scaling~\cite{Feynman:1969ej}. According to measurements the scaling hypothesis holds at least up to $7\,$TeV~\cite{Pare:1989mr,Adriani:2015iwv}. \figu{pi0-scaling} shows the scaling behavior of \sibyll{} in comparison with the data, which is compatible within the errors of the data.

In the calculation of the atmospheric fluxes the production spectra enter in a weighted form (\eg{} \equ{z-factor}), emphasizing the very forward region. At the same time the primary flux covers center-of-mass energies from $10\,$GeV to $100\,$TeV. The production spectra for \sibyll{} with a weight for a typical cosmic ray spectrum are shown in \figu{sib-scaling}. From the presence of scaling violations in the central region due to multi-parton interactions a slight softening of the spectra with energy is expected~\cite{Ostapchenko:2016ytp}. With the new remnant model \sibyll{} initially showed a hardening of the meson spectra with interaction energy\footnote{intermediate version \sibyll~2.3}. This behavior was found to originate from the splitting of leading di-quarks, originally introduced to improve the leading flavor ratios~\cite{Riehn:2017mfm}. With \sibyll{-2.3c} scaling is approximately fulfilled.

The effects of these changes on the atmospheric flux of muons and the
charge ratio of muons are discussed in Sect.~\ref{ssec:charge_ratio}.

\subsection{Charm production}
\label{ssec:sib23_charm}

\subsubsection{Charm model}
\label{subsec:charm-model}
\begin{figure}
  \includegraphics[width=\columnwidth]{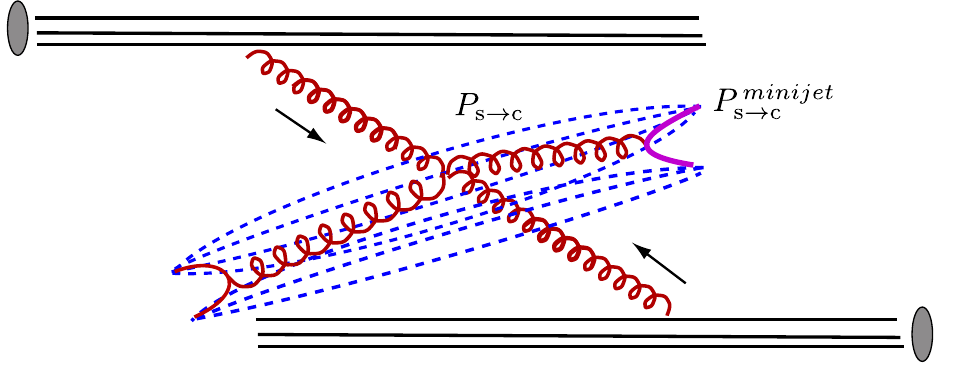}
  \caption{\label{fig:sib-minijet-charm} String configuration for
    minijets in \sibyll{}. }
\end{figure}

\begin{figure}
  \includegraphics[width=0.8\columnwidth]{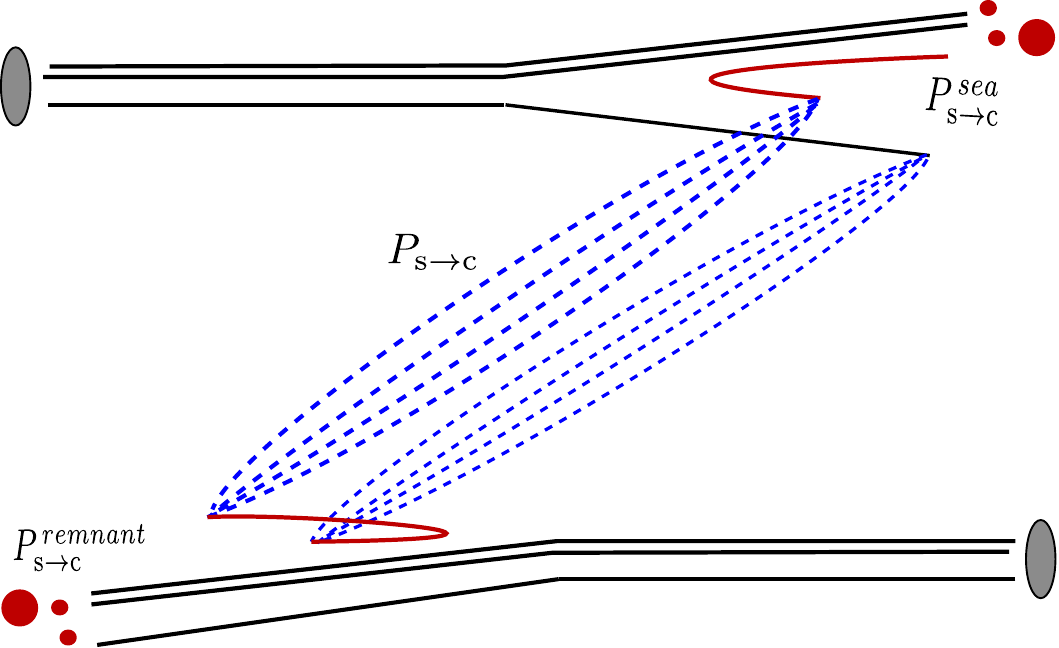}
  \caption{\label{fig:sib-leading-charm} String configuration with remnant excitation. Note that the flavor content of the remnant can be different from the initial hadron in the case of remnant excitation.}
\end{figure}

Despite their low production yield, charmed hadrons play an important role for the prompt flux of high energy leptons in the atmosphere (Sect.~\ref{ssec:relevant_hadrons} and Sect.~\ref{ssec:prompt_flux_discussion}).  We have therefore included them in the model.
The production of heavy flavors in hadron interactions is qualitatively different from the production of light flavors. While the light u,d and s quarks are abundantly produced in soft fragmentation processes, the production of heavy quarks due to the large mass ($m_{\rm c\bar{c}}\approx \mathcal{O}(2$\,GeV$)$) is well above the soft scale. Production of charmed hadrons can therefore be expected to be dominated by hard (perturbative) processes as in \figu{sib-minijet-charm}. On the other hand, measurements of leading charm production ($x_{\rm F} > 0.6$) at low energy~\cite{Aitala:1996hf,Garcia:2001xj} indicate that there is a soft (non-perturbative) component~\cite{frixione1994charm,Frixione:1997ma}. Associated production of
charm, (e.g. production of $\Lambda_c + D^0$) is illustrated in \figu{sib-leading-charm}.  

The model of charm quark production in \sibyll{-2.3c} uses the family connection between strange and charmed hadrons, exchanging s with c quarks in the fragmentation step~\cite{Ahn:2011wt}. The total rate of charm production in this model is set by the probability $P_{\rm s \to c}$
for the replacement of a strange quark by a charm quark. At energies beyond a TeV, when the mass of the charm quark becomes negligible in comparison with the scale of the parton interactions, the difference between the light flavors and charm vanishes and the energy dependence is given by the minijet cross section. In this energy regime a fixed charm to strange rate $P_{\rm s \to c}$ is appropriate. In the threshold region charm production increases much more rapidly with the \cm{} energy. To account for this threshold the charm rate in minijets decreases as
\begin{equation}
P_{\rm s \to c} ~=~ P_{0, \,\rm s \to c} \cdot \exp{ [ -m_{\rm c,eff} / \sqrt{\hat{s}} \, ] } \ , \nonumber
\end{equation} 
where $\sqrt{\hat{s}}$ is the \cm{} energy of the frame of the partons and $m_{\rm  eff}=10\,$GeV is the effective charm mass. The fragmentation function used for charm quarks is the SLAC/Peterson parameterization with $\epsilon=2.0$~\cite{Peterson:1982ak}. The transverse momentum of the charm pair in the string is taken from an exponential transverse mass distribution with energy dependent mean
\begin{equation}
\langle m_{\rm T} \rangle(s) ~=~ p_{\rm T,0} + ( 0.3\,\mathrm{GeV}) \, \log_{10}{(\sqrt{s}/ \, 30\,\mathrm{GeV})} \ , \nonumber
\end{equation}
where $p_{\rm T,0}$ is $0.3\,$GeV for charmed mesons and $0.5\,$GeV
for charmed baryons.

\begin{figure}
  \includegraphics[width=\columnwidth]{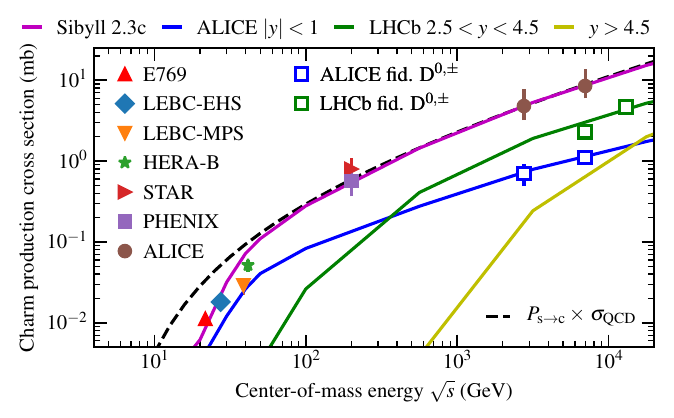}
  \caption{\label{fig:sib-charmprod}Inclusive cross section of the
    production of charmed quarks. Data below $100$\,GeV are from
    fixed-target
 experiments~\cite{Alves:1996rz,AguilarBenitez:1988sb,Ammar:1988ta,Zoccoli:2005yn,Lourenco:2006vw}.
    Intermediate and high energy data are from the RHIC and the
    LHC~\cite{Adamczyk:2012af,Adare:2010de,Abelev:2012vra,ALICE:2011aa,Aaij:2013mga,Aaij:2015bpa}.
    Apart from the full cross section (magenta line), cross sections
    for different phase space cuts of the LHC experiments are
    shown. The blue line corresponds to the central phase space
    covered in ALICE ($|y|<0.5$), while the green line corresponds to
    the more forward coverage of LHCb ($2.5<y<4.5$). In yellow the
    contribution from phase space not covered by any experiment is
    shown. The dashed line represents the cross section for hard
    parton scattering scaled to represent charm production in the
    model.}
\end{figure}
\begin{figure}
  \includegraphics[width=\columnwidth]{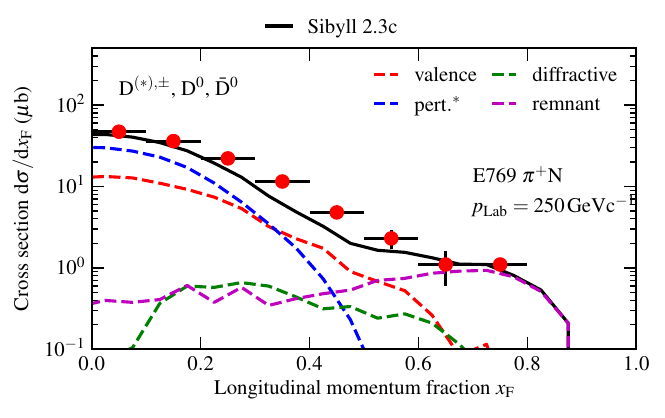}
  \caption{\label{fig:sib-e769-charm}Feynman-$x$ spectrum of charmed mesons in $\pi$-nucleus interactions. Data are from E769~\cite{Alves:1996qz}, taken with beam momentum of $250\,$GeV$/$c. Dashed lines are contributions from individual processes in \sibyll{}. The ''remnant'' contribution arises from the exchange of a quark with the soft strings (see Fig.~\ref{fig:sib-leading-charm}).}
\end{figure}

\begin{figure}
  \includegraphics[width=\columnwidth]{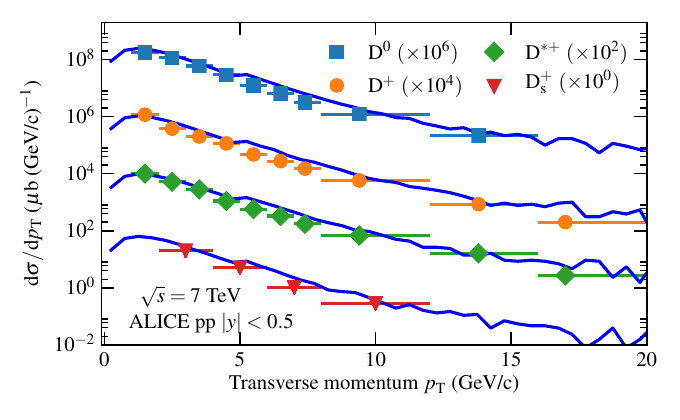}
  \caption{\label{fig:sib-alice-charm}Spectrum of the transverse
    momentum of D-mesons. Measurement from ALICE is taken in the
    central phase space $|y|<0.5$ at
    $\sqrt{s}=7\,$TeV~\cite{ALICE:2011aa}. Relative abundance of
    D-mesons, vector resonances ($\mathrm{D}^\star$) and D-mesons
    containing a strange quark ($\mathrm{D}_{\rm s}$) is determined by the
    parameters in the strange quark sector. The line is the prediction from \sibyll{-2.3c}. }
\end{figure}

\begin{figure}
  \includegraphics[width=\columnwidth]{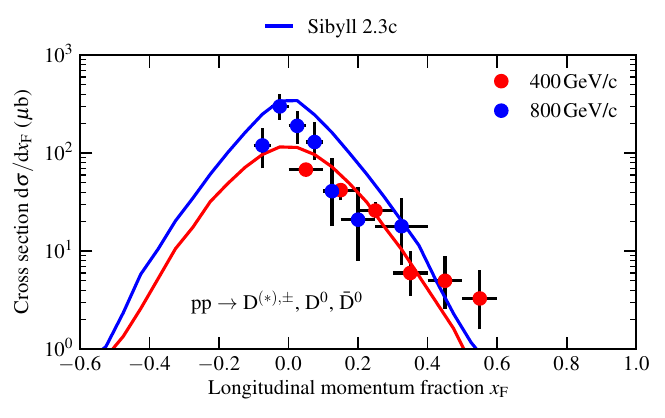}
  \caption{\label{fig:sib-lebc-charm}Feynman-$x$ spectrum of charmed
    mesons in pp interactions. Data are from LEBC experiment~\cite{AguilarBenitez:1987rc,Ammar:1988ta}, recorded at beam momenta of $400$ and $800\,$GeV$/$c. The lines are the predictions from \sibyll{-2.3c}.}
\end{figure}

\begin{figure*}
  \includegraphics[width=\columnwidth]{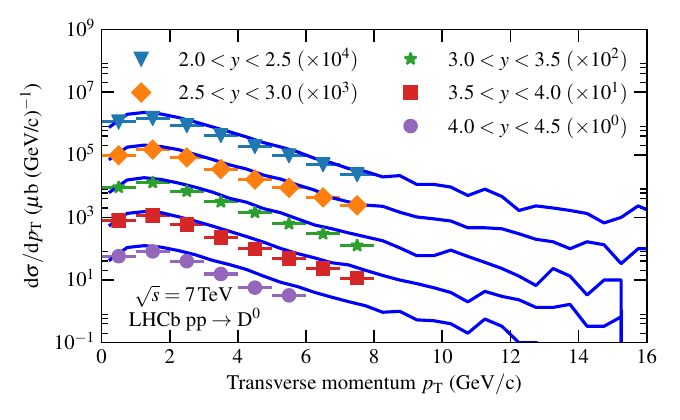}
  \hfill
  \includegraphics[width=\columnwidth]{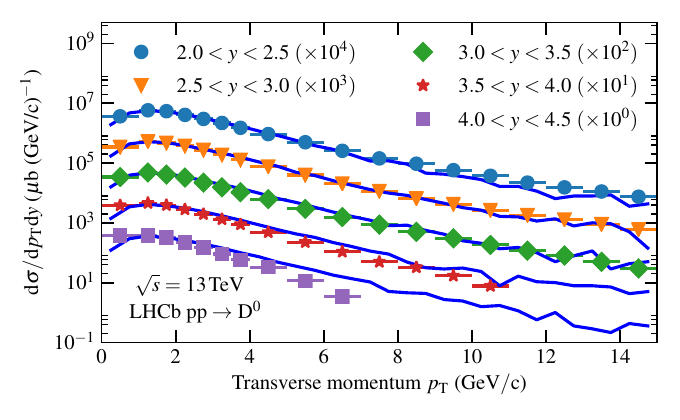}
  \caption{\label{fig:sib-lhcb} $p_{\rm T}$-spectrum of neutral D-mesons in LHCb at $7$ and $13\,$TeV \cm{} energy~\cite{Aaij:2013mga,Aaij:2015bpa} compared to \sibyll{~2.3c}}
\end{figure*}
The comparison of the inclusive ${\rm c\bar{c}}$ production cross section of the model with measurements in a wide range of energies is presented in \figu{sib-charmprod}. Apart from the full inclusive cross section, the figure also shows the cross section in central (ALICE), next-to-central (LHCb) and forward phase space. The model correctly describes the growth with energy in the most central phase space, but not as well in the next-to-central region. The different behavior in the threshold region and at large energies can be seen by comparing charm production in the model with the scaled minijet cross section $\sigma_{\rm QCD}$ in \figu{sib-charmprod}.

While this phenomenological model of charm production is sufficient to describe the total charm yield, it has difficulties to describe differential spectra, like the transverse momentum spectra of charmed hadrons measured by LHCb~\cite{Aaij:2013mga,Aaij:2015bpa} and
ALICE~\cite{ALICE:2011aa}, as it neglects the details of hard scattering. To account for the dominant contribution from hard scattering without altering the minijet model at the basis of \sibyll{}, the production of charm quarks is focused towards the string ends by invoking a separate $P^{\rm \, Minijet}_{\rm s \to c}$ for the gluon splitting (${\rm g \to q\bar{q}}$), as illustrated in \figu{sib-minijet-charm}.

The nature of the production of charmed hadrons from soft interactions is not well understood. The spectra of leading D-mesons in p-, $\pi$-, and K-nucleus interactions measured at the E769 experiment, together with the asymmetry between $\Lambda_{\rm c}$ and its antiparticle observed in pp scattering by the SELEX collaboration, indicate the presence of a non-perturbative, soft component~\cite{Alves:1996rz,Garcia:2001xj}. Irrespective of the exact
origin of the leading charm production, the hard minijet component, representing perturbative interactions in \sibyll{} is not sufficient to describe the low energy data. In \sibyll{} soft interactions include soft minijets, the scattering of the valence quarks, the formation of beam remnants and diffraction dissociation. The soft minijets in the model derive from soft gluon scattering which has a steep longitudinal momentum distribution, resulting again in mostly central production. In contrast, the momentum spectrum for valence quarks and the remnant is much harder, so the hadron spectra at large $x_{\rm F}$ are determined by these processes. \figu{sib-e769-charm} shows how the different processes form the D-meson spectrum in pion-nucleus interactions at the E769 experiment. While minijets below the hard scale in the framework of the model are referred to as soft, they are still within the range of perturbative calculations. Correspondingly contributions from soft and hard minijets in the model are shown together as $\mathrm{pert.}^*$. Truly non-perturbative contributions are D mesons from the fragmentation of the valence quarks, the diffractive contribution and the remnant (all labeled accordingly). In addition to the rate of charm from string fragmentation and in minijets, the model generates charm in the splitting of the soft gluons ($P^{\rm Sea}_{\rm s \to c}$) and as an intrinsic charm content in the remnant ($P^{\rm Remnant}_{\rm s \to c}$). The sources for the leading charm production in the soft processes are sketched in \figu{sib-leading-charm}.

The production of D-mesons in the central region $|y|<0.5$ as a function of the transverse momentum is shown in \figu{sib-alice-charm}, compared with the measurement from ALICE at $7\,$TeV \cm{} energy~\cite{ALICE:2011aa}. Since the central region is dominated by hard scattering, the charm rate for the hard minijets ($P^{\rm \, Minijet}_{\rm s \to c}$) is determined from these data. The difficulty in describing the growth rate of the total yield in the LHCb phase space in \figu{sib-charmprod} also appears in the differential spectra shown in \figu{sib-lhcb}. The yield at $7\,$TeV is overestimated, and the shape of the $p_T$ spectra deviates at higher $p_{\rm T}$ values and at large rapidities. The same trend continues at $13\,$TeV, although the model predicts the correct overall yield. These discrepancies are probably connected to the approximation of the perturbative charm production with the hard scattering cross section and the limitations of the simple minijet model.

The parameters of soft charm production are determined by the low energy pion-nucleon and proton-proton data shown in \figu{sib-lebc-charm} and \figu{sib-e769-charm}. It is interesting to note that in order to describe the forward production at E769, charm production in the central strings and in soft gluons is sufficient. No charm production in the hadronization of the remnant is required. The numerical values of the parameters of the charm model are summarized in Tab.~\ref{tab:charm-param}. An overview of the available measurements that have been used for the determination of free model parameters is in Table \ref{tab:charm-data}.

\begin{table}
  \caption{Table of the free model parameters that control the probabilities for charmed quarks production in different processes. As described in more detail in the text, the effective rates can be attenuated by additional factors. 
    \label{tab:charm-param}}
  \begin{center}
    \renewcommand{\arraystretch}{1.5}
    \begin{tabular}{cc}
      \hline
      \hline
      parameter & value \\
      \hline
      perturbative & \\
      $P^{\rm minijet}_{\rm s \to c}$ & 0.08 \\
      non-perturbative  & \\
      $P^{\rm soft}_{\rm s \to c}$ & 0.004 \\
      $P^{\rm sea}_{\rm s \to c}$ & 0.002 \\
      $P^{\rm remnant}_{\rm s \to c}$ & 0.0 \\
      $P^{\rm string}_{\rm s \to c}$ & 0.004 \\
      \hline
    \end{tabular}
  \end{center}
\end{table}

\begin{table*}
  \caption{Experiments that collected data on charm production including the corresponding projectile-target configuration and the accessible longitudinal phase space. These data have been used for model development and parameter estimation. \label{tab:charm-data}}
  \begin{center}
    \renewcommand{\arraystretch}{1.5}
    \begin{tabular}{ccccccc}
      \hline
      \hline
      Name & $P_{\rm Lab}$ (GeV) & $\sqrt{s}$ (GeV) & $x_{\rm F}$ spectrum & $x_{\rm F}$ coverage & Beam config. & Ref. \\
      \hline
      E-769 & $250$ & $22$ & yes & $-0.1<x_{\rm F} < 0.8$ & p-Nuc & \cite{Alves:1996rz,Alves:1996qz}\\
      EHS & $400$ & $27.4$ & yes & $0<x_{\rm F} < 0.6$ & p-p & \cite{AguilarBenitez:1988sb,AguilarBenitez:1987rc}\\
      MPS & $800$ & $39$ & yes & $-0.1<x_{\rm F} < 0.4$ & p-p & \cite{Ammar:1988ta}\\
      HERA-B & $920$ & $42$ & no & $-0.1<x_{\rm F} < 0.05$ & p-Nuc & \cite{Zoccoli:2005yn}\\
      STAR & $21\,$TeV & $200$ & no & $-0.03<x_{\rm F} < 0.03$ & p-p & \cite{Adamczyk:2012af}\\
      PHENIX & $21\,$TeV & $200$ & no & $-0.003<x_{\rm F} < 0.003$ & p-p & \cite{Adare:2010de}\\
      ALICE & $4\,$PeV & $2.76\,$TeV & no & $-0.005<x_{\rm F} < 0.005$ & p-p & \cite{Abelev:2012vra}\\
      & $26\,$PeV & $7\,$TeV & no & $-0.004<x_{\rm F} < 0.004$ & p-p & \cite{ALICE:2011aa}\\
      LHCb & $26\,$PeV & $7\,$TeV & no & $0.002<x_{\rm F} < 0.1$ & p-p & \cite{Aaij:2013mga}\\
      & $90\,$PeV & $13\,$TeV & no & $0.002<x_{\rm F} < 0.1$ & p-p & \cite{Aaij:2015bpa}\\
      \hline
    \end{tabular}
  \end{center}
\end{table*}

\subsubsection{Comparison with other models \& discussion}
\label{ssec:comp_with_other_charm}
\begin{figure}
  \includegraphics[width=\columnwidth]{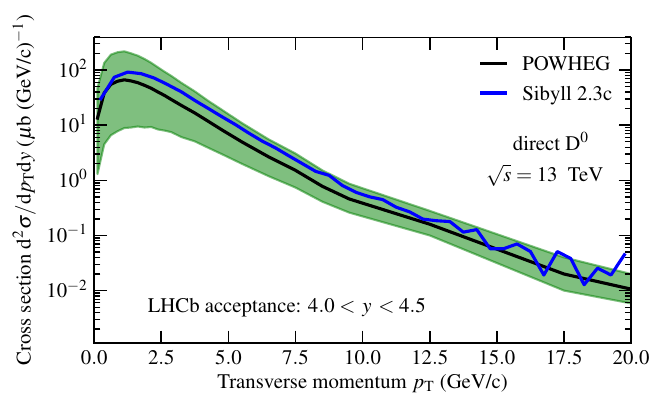}
  \caption{\label{fig:sib-charm-pqcd} $p_{\rm T}$-spectrum of neutral
    D-mesons at $\sqrt{s}=13$\,TeV. Compared are \sibyll{} and a NLO 
    QCD calculation (POWHEG)~\cite{Gauld:2015yia}. The band of the theory
    calculation corresponds to charm quark mass and factorization
    scale uncertainty. The central phase space covered by the LHCb
    acceptance is dominated by perturbative processes, such that the
    full prediction by \sibyll{} can be compared to the pQCD
    calculation.}
\end{figure}

\figu{sib-charm-pqcd} compares the transverse momentum spectrum of neutral D-mesons in \sibyll{} and a more fundamental next-to-leading order (NLO) QCD calculation based on POWHEG in the LHCb phase space ~\cite{Gauld:2015yia}. The spectra match nicely within the theory uncertainty, indicating that our phenomenological charm model is sufficient to describe the transverse momenta for perturbatively produced charm. As mentioned previously, the hard scale
in the model is not equivalent to the scale in pQCD calculations. However, in the (still) central phase space covered by LHCb, the comparison is valid between pQCD calculations and the full prediction from \sibyll{}.

\begin{figure}
  \includegraphics[width=\columnwidth]{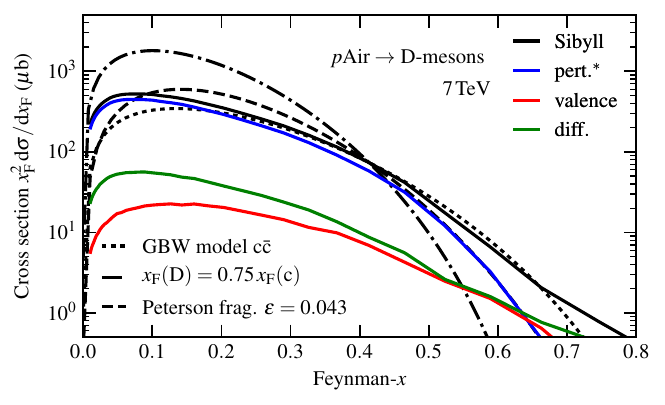}
  \caption{\label{fig:sib-charm-mrs} Longitudinal momentum spectrum of
    D-mesons at $\sqrt{s}=13$\,TeV. \sibyll{} is compared with a
    calculation in perturbative QCD using extensions to small-$x$
    (GBW)~\cite{Martin:2003us,GolecBiernat:1998js,GolecBiernat:1999qd}. GBW
    model is scaled to match \sibyll{} at $x_{\rm F}\sim 0.4$. While
    the yields are different, the shape of the spectrum predicted for
    the perturbative component is comparable.}
\end{figure}

For the inclusive lepton fluxes, the particle production in longitudinal phase space is paramount. However, calculations that extend into the region of large Feynman-$x$ / rapidity are difficult as they include scatterings small-$x$ ($y\sim |x_1-x_2 |$, $x_1$, $x_2$ are the momentum fractions of the partons) and typically require additional assumptions. \figu{sib-charm-mrs} shows the $x_{\rm F}$ spectrum of D-mesons at $13\,$TeV, comparing \sibyll{} and a small-$x$ color-dipole calculation including saturation effects (GBW)~\cite{Martin:2003us,GolecBiernat:1998js,GolecBiernat:1999qd}. In the figure the GBW calculation is re-scaled to match our spectrum at $x_{\rm  F}\approx 0.4$. Similar to the POWHEG calculation (\figu{sib-charm-pqcd}), the $\mathrm{pert.}^*$ component in \sibyll{} and the GBW calculation match well, in particular when the SLAC/Peterson fragmentation function is applied. In our model, the additional non-perturbative components (valence quark interactions, remnant and diffractive processes) start to dominate the spectrum spectrum around $x_{\rm F}=0.7$. Note that a weight of $x_{\rm F}^2$ is applied in the figure, which is the appropriate weight to study prompt leptons which originate from interaction energies above the cosmic ray knee.

\begin{figure}
  \includegraphics[width=\columnwidth]{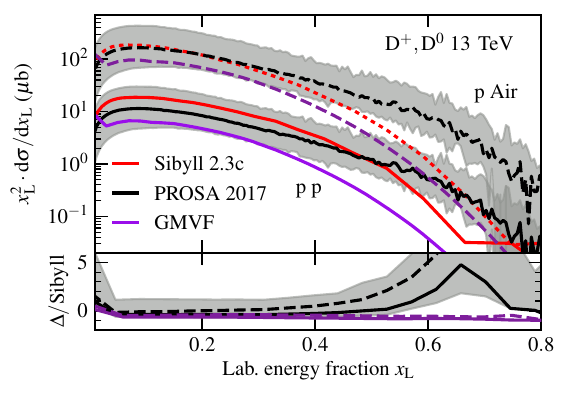}
  \caption{\label{fig:sib-charm-mvg} Comparison of the energy spectra of charmed mesons between \sibyll{} (red lines) and NLO QCD calculations (PROSA 2017~\cite{Garzelli:2016xmx} (black lines) and GMVF~\cite{Benzke:2017yjn} (purple lines)). Energy is evaluated in the lab.\ frame. The lower panel shows the difference relative to \sibyll{}. The grey band represents the theoretical uncertainty in the PROSA calculation (factorization and renormalization scale). Calculations for pp interactions are shown by solid lines and those for p-air with are dashed or dotted. The NLO calculations for an Air target are scaled up according to $14.5 \times $pp. }
\end{figure}

\figu{sib-charm-mvg} shows a comparison to more recent NLO QCD calculations, PROSA 2017~\cite{Garzelli:2016xmx} and GMVF~\cite{Benzke:2017yjn}. The PROSA calculation takes into account charm production from hard scattering with leading logarithmic corrections from parton showers. Non-perturbative effects from hadronization and leading particle effects are included through PYTHIA~8~\cite{Sjostrand:2015cx}. In the GMVF calculation hadronization is included via fragmentation functions. In contrast to GBW, PROSA and GMVF do not include a specific model for the small-$x$ region of the parton distribution functions, instead they are fixed by and extrapolated from HERA and LHCb data~\cite{Zenaiev:2015rfa}. The transition from proton-proton to proton-air interactions is simplified through a superposition model $\sigma_{c\bar{c}, {\rm p-Air}} = 14.5\times\sigma_{c\bar{c}, {\rm pp}}$. Within the theory uncertainty (see bands in the figure) which contains contributions from the variation of the renormalization and factorization scales, the $\xl$-spectra from PROSA and \sibyll{} are compatible. While the difference in the normalization between GMVF and \sibyll{} is larger than for PROSA, the shape of the spectrum is more similar.
Note that due to the additional scattering centers in the target nucleus, it is expected that the production spectra are attenuated at large $\xl$. Since the superposition ansatz is used for PROSA and GMVF, the difference with \sibyll{} increases for proton-air.

\begin{figure}
  \includegraphics[width=\columnwidth]{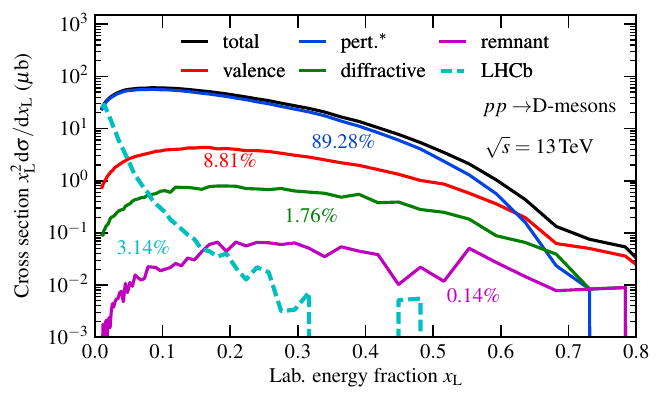}
  \caption{\label{fig:sib-xl-spec} Weighted energy spectrum of
    D-mesons at $\sqrt{s}=13$\,TeV. The energy of the particles is
    taken in the lab.\ frame and expressed relative to the beam
    energy. Contributions by different phase space intervals are
    shown. The contribution of the phase space covered by LHCb in
    events that also trigger LHCb is around $3\,$\%.}
\end{figure}

Finally, \figu{sib-xl-spec} shows the spectrum of D-mesons as a function of the energy fraction in the lab.\ frame, also weighted by $\xl^2$, and broken down into different production processes. Their relative contribution to the spectrum weighted moment ($Z$-factor) is printed as a percentage value. While the soft processes in \sibyll{} (valence, diffractive and remnant) are important to describe the shape of the longitudinal spectra at low energy, their contribution to the $Z$-factor at high energy is only of the order of $10\,$\%. Minijet charm production, in contrast, sums up to $89\,$\%, resulting in a seemingly well determined the $Z$-factor. Despite the fact that the model only approximates perturbative charm production and shows slight deviations in the next-to-central phase space, the constraint on the $Z$-factor will be not quite as strong. Quantitatively, while $89\,$\% of charm is produced in perturbative processes in the model, just $3\,$\% are covered and constrained by collider measurements (LHCb). Under the assumption that the phase space extrapolation of the perturbative processes is well understood, this leaves about $10\,\%$ of charm
production (valence, diffractive and remnant processes) unconstrained by the LHC.

With current LHC detectors there is little hope to directly constrain the model at higher energies, since the LHCb detector is already hitting the technological limit concerning particle identification at small angles.  Theoretically the model may be constrained by improving the perturbative model, in order to reduce the theoretical uncertainties. Additional input is expected from (next generation) large volume neutrino telescopes that can be sensitive to the prompt neutrino flux through a measurement of certain ratios (see Sects. \ref{ssec:neutrino_ratios} and \ref{ssec:vh_ratios}).

At lower energies on the other hand the fixed-target configuration of LHCb~\cite{Lansberg:2012wj,Ulrich:2015uwb} may provide valuable input. While the boost from \cm{} to the lab.\ frame worsens the situation on the beam side, the target region will be shifted into the acceptance of LHCb. Measuring the production of D-mesons in fixed-target proton-proton, or ideally oxygen-proton, collisions with LHC beams ($6.5\,$TeV) could determine the $x_{\rm L}$-spectrum in the intermediate energy range of $\sqrt{s}\sim 110\,$GeV. For comparison, the current highest energy measurement of the longitudinal spectrum of D-mesons is at $40\,$GeV.

\section{Impact of the improved \sibyll{-2.3c} on inclusive flux calculations}
\label{sec:impact}

In this last section, we challenge the new model against its predecessor \sibyll{-2.1} and computations using other recent hadronic interactions models. The model {\sc EPOS-LHC} \cite{Pierog:2013ria} is currently very successful in describing cosmic ray air-shower observations and minimum-bias collider data at the same time. A representative model using the Quark-Gluon-String framework (an analogous approach to the Dual Parton Model) is {\sc QGSJet-II-04} \cite{Ostapchenko:2010vb}. \revise{The \dpmjet{-III-19.1} \cite{Fedynitch:2015kcn} is an updated version of \dpmjet{-III-3.0.6} event generator. It better describes LHC minimum-bias data and contains a more sophisticated approach for cascade calculations. All three} models have been revisited after the launch of the LHC and have been adjusted to similarly recent data as \sibyll{-2.3c}. We can easily swap the interaction model in our \mceq{} calculations and keep all other parameters equal to the \sibyll{-2.3c} calculations when doing this.

In addition, we compare to the very detailed predictions of atmospheric neutrinos from the two state-of-the-art calculations; {\it HKKMS 2015} \cite{Honda:2015fha} is based on a fully 3-dimensional geometry (3D) at energies below 30 GeV and on a 1D Monte-Carlo calculation at higher energies. The other reference is known as the {\it Bartol} calculation \cite{Barr:2004br} and consists of a 3D part at energies below 10 GeV and on a 1D Monte Carlo above that. The history of the HKKM and the Bartol calculations started over 20 years ago and played an important role in the era of the Super-Kamiokande detector and the discovery of neutrino oscillations. 

Both calculations use similar technology and the parameterization of the cosmic ray flux from \cite{Gaisser:2002jj}. In HKKMS, hadronic interactions are modeled with an effective particle Monte Carlo that is based on inclusive parameterizations of secondary particle spectra from the \dpmjet{-III 3.0.5} event generator \cite{dpmjetIII}. Since without modifications \dpmjet{} does not describe atmospheric muon fluxes to the required precision, corrections to particle production have been applied \cite{Honda:2006qj} derived from muon measurements. These corrections do not have a microscopic explanation and depend on the choice of the primary cosmic ray parameterization \revise{(see discussion in Sect.~\ref{sec:prim_flux_dependence_neutrinos})}.

The Bartol calculation uses the {\sc Target-2.1a} Monte Carlo particle generator, an effective method that generates secondary particles according to parameterizations of fixed target data, conserving some physical quantities like energy and momentum. Its advantage is that in the energy range covered by fixed target data, the resulting spectra will converge to these measurements for a large number of Monte Carlo samples. However, the lack of a detailed microphysical model limits the extrapolation capabilities into a phase space without measurements. An important example of such an ``extrapolation problem'' is the associated production of K$^+$ in the process ${\rm p} + {\rm air} \to \Lambda + {\rm K}^+$. The authors of \cite{Barr:2004br} seem to have made an optimistic choice for this particular process that is somewhat in tension with the current (microscopic) hadronic interaction models. Since there are no fixed target results on K$^\pm$ production at projectile momenta above 400 GeV, this choice can not be constrained by data and impacts the uncertainties of the present calculations.

In addition, we compare to other 1D muon \cite{Kochanov:2008pt} and neutrino flux \cite{Sinegovskaya:2014pia} calculations that rely on numerical methods \cite{Naumov:1984jfv,Naumov:1993yr}. \revise{The hadronic interactions are parameterized inclusive secondary particle spectra from measurements or event generators, very similar to the possibilities in \mceq{}}.

In the following sub-sections we explicitly do not aim to discuss the uncertainties of the calculations and leave this topic to a follow-up publication. For this reason we avoid  making comparisons with inclusive flux measurements, in particular because the fluxes notably depend on the choice of the cosmic ray spectrum. The reference models were extensively compared to data and can act as a reference point. The H3a nucleon flux (Sect.~\ref{sec:primary_flux_H3a} is used as the primary cosmic ray flux throughout all calculations \revise{or if not otherwise mentioned. Note, that this primary flux model is mainly derived from high-energy data. A good description of fluxes at low energies ($<100$ GeV) can therefore not be expected.} It is important to keep in mind that the \mceq{} based calculations are 1D calculations without accounting for the geomagnetic cutoff, solar modulation effects or Earth propagation for up-going fluxes. These approximations result in up-/down- and azimuth symmetric fluxes by design and do not depend on the choice of a ``detector'' location. As discussed in the literature concerning the 3D modeling, these approximations are good at energies $E_{\rm lepton} > 5$ GeV. The characterization of fluxes below 5 GeV has been a very active topic in the past (see for example the review \cite{Gaisser:2002jj} or the references in \cite{Barr:2004br,Honda:2015fha}) and goes beyond the scope of this work. The calculations in the following sections have been made with \mceq{} version 1.0.8.

\subsection{Fluxes}
\revise{In this section, we focus on the discussion of the spectra of inclusive leptons. The angular distributions are presented in the next section.}
\subsubsection{\revise{Fractional contributions}}
\begin{figure}
  \includegraphics[width=\columnwidth]{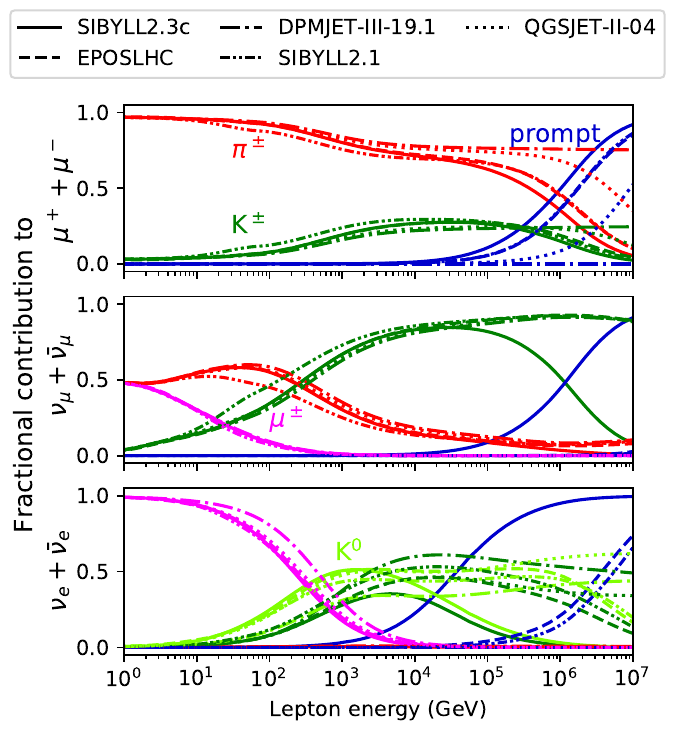}
  \caption{\label{fig:meson_fractions} Fractions of hadrons decaying into atmospheric leptons. The muon flux (upper panel) is calculated with the indicated interaction models for $\theta=0^\circ$. The neutrino fluxes (lower panels) show the results for zenith averaged fluxes.}
\end{figure}
The break-down of the different hadron species decaying into atmospheric muons and neutrinos is shown in \figu{meson_fractions} for different choices of the hadronic interaction model. In the energy range where conventional leptons dominate, the recent models \sibyll{-2.3c}, \eposlhc{} and \qgsjet{} have almost identical behavior. \revise{\dpmjet{} has the smallest kaon contribution, while \sibyll{-2.1} the kaon component is more abundant. The importance of the kaon component can be studied by comparing these two models.}

\revise{\figu{meson_fractions} shows the crossover between conventional components and prompt leptons (see discussion in Sec.~\ref{ssec:angular_dist}). The small prompt contribution at very high energies for \sibyll{-2.1} and \dpmjet{-III-19.1} originates from rare decays and is in practice not relevant. Note that \dpmjet{-III-19.1} produces charmed particles by default that would result in a prompt flux. This is a natural effect of the updated parton distribution functions and appears as a part of the hard component. However, the charm production has not been explicitly studied, as we have performed here in the case of \sibyll{}, and hence it will disagree with data. The prompt component from \dpmjet{-III-19.1} is not recommended for any practical purpose and has been disabled in \mceq{} for the following sections. }
\subsubsection{Muons}
\begin{figure*}
  \includegraphics[width=\textwidth]{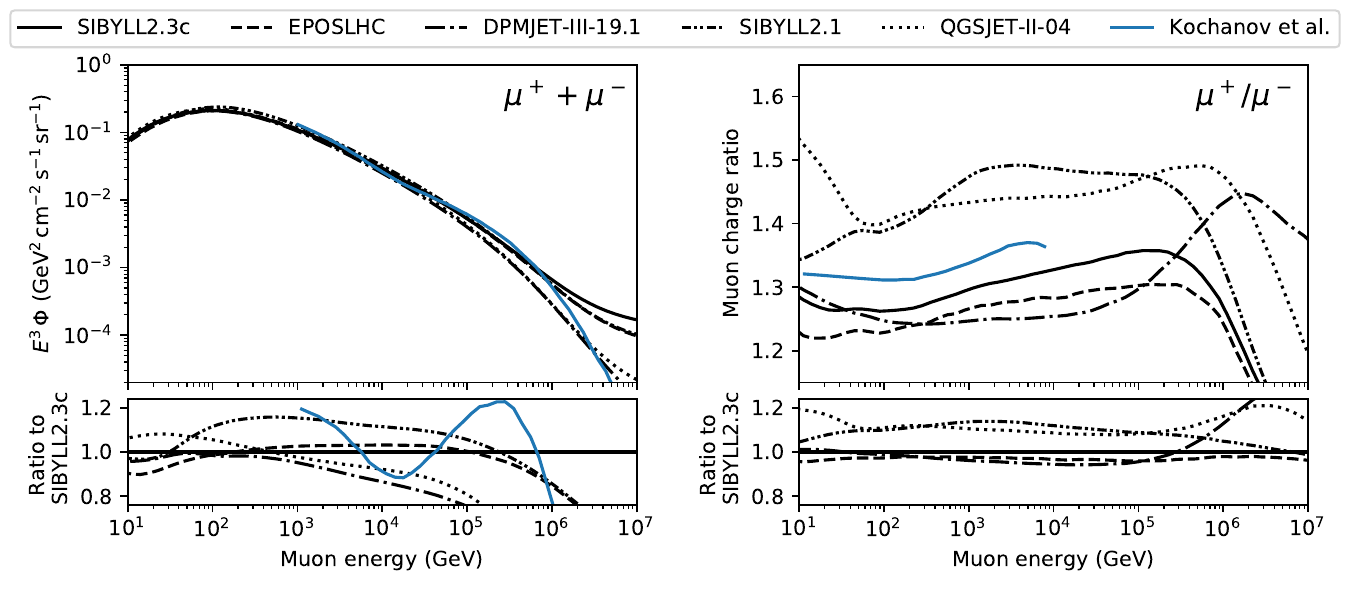}
  \caption{\label{fig:vertical_muon_fluxes} The atmospheric muon flux and the charge ratio. The \mceq{} fluxes for different interaction models are calculated for the zenith angle $\theta = 0^\circ$ and the H3a primary spectrum. For the {\it Kochanov et al.} \cite{Kochanov:2008pt} curves the authors used the KM parameterization for hadronic interactions and the Zatsepin-Sokolskaya primary spectrum.}
\end{figure*} 

In \figu{vertical_muon_fluxes} we compare muon flux predictions using the different hadronic models in \mceq{} and another numerical calculation. The spread is within 10\% for the post-LHC interaction models and increases to 20\% when including \sibyll{-2.1}.  As the muon charge ratio in the right panel suggests, this has to do with an enhanced production of K$^+$ that originates from a program artifact in the old version.  These K$^+$ are copiously overproduced when diffraction occurs on nuclear targets. A correction of this behavior leads to a flat and rather small charge ratio. As expected from the explanations in Section \ref{ssec:charge_ratio}, \sibyll{-2.3c} and \eposlhc{} predict an increase of the charge ratio at higher energies. The ``wavy'' behavior of the {\it Kochanov et al.} calculation is related to its primary spectrum and not the hadronic model, that assumes scaling at high energies.

In summary, the muon fluxes seem well constrained since they mostly depend on the modeling of the secondary pion production, where no such associated production channel as for K$^+$ exists.
The charge ratio is more sensitive to the details of forward kaons, as discussed in \cite{Gaisser:2012em} (see also \cite{Agafonova:2014mzx}). While it seems constrained within 15\% by the model predictions, this range exceeds typical experimental errors of a few percent. Also, the cosmic ray composition impacts the charge ratio with a similar magnitude.  

\subsubsection{Neutrinos}
\begin{figure*}
  \includegraphics[width=\textwidth]{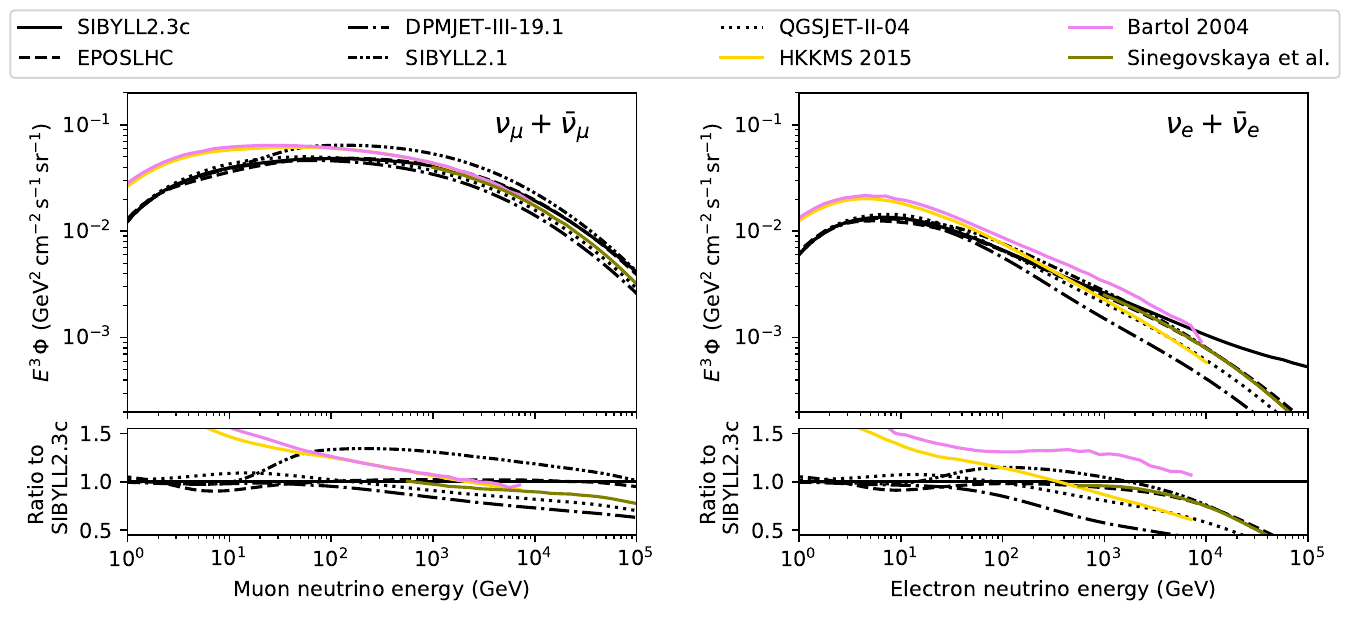}
  \caption{\label{fig:zenith_averaged_neutrinos} Atmospheric muon and neutrino fluxes averaged over the zenith angle. The HKKMS and Bartol curves are from computations for Kamioka site and Solar Flux minimum. At energies above a few GeV the dependence on the detector location and solar modulation diminishes. The curves by {\it Sinegovskaya et al.} are computed for the similar choice of models as in our case, namely an H3a primary spectrum and \qgsjet{-II-03} interaction model. }
\end{figure*}

The muon neutrino fluxes (left panel of \figu{zenith_averaged_neutrinos}) calculated with the recent interaction models are very similar. As mentioned above, the K$^+$ issue in \sibyll{-2.1} is responsible for the deviation of $\sim 30\%$ from the other models and is indeed unphysical. The disagreement of the neutrino spectral index between our calculations and HKKMS or Bartol is caused in part by the different choice of the primary flux \revise{(see Sect.~\ref{sec:prim_flux_dependence_neutrinos}). Differences may also come \revise{from} the treatment of decays, which in case of \mceq{}} are simulated with the \pythia{-8} Monte Carlo \cite{Sjostrand:2015cx} using methods that preserve high accuracy throughout all steps of the calculations. We find that the calculation by {\it Sinegovskaya et al.} \cite{Sinegovskaya:2014pia} \revise{produce a few percent higher flux} using exactly the same set of models (olive curves in \figu{zenith_averaged_neutrinos}), consistent with the findings by \cite{Morozova:2017fof}. For Bartol, the higher number of electron neutrinos \revise{is} the result of a higher kaon abundance in associated K$^+$ production in {\sc Target-2.1a}.  In the case of HKKMS \revise{sizable deviations are expected, since the interaction model has been modified to match muon flux and charge ratio measurements.} However, we can not yet verify these conjectures at the level of hadronic interactions. In the case of the angular distribution of inclusive leptons, \revise{differences in geometry and computational efficiency may also contribute to slightly different results. At energies below 80 GeV the calculations sizably diverge. As discussed below in Sect.~\ref{sec:prim_flux_dependence_neutrinos}, the spectral mismatch comes from the choice of the primary flux model.} \footnote{\revise{Note, that the \mceq{} results changed at the lowest energies in the final version of the code. In the previous versions, too many nucleons have been recorded in the tables of the low energy hadronic interaction model (\dpmjet{-III-17.1}).}}

\subsubsection{Prompt fluxes from atmospheric charm}
\label{ssec:prompt_flux_discussion}
\begin{figure}
  \includegraphics[width=\columnwidth]{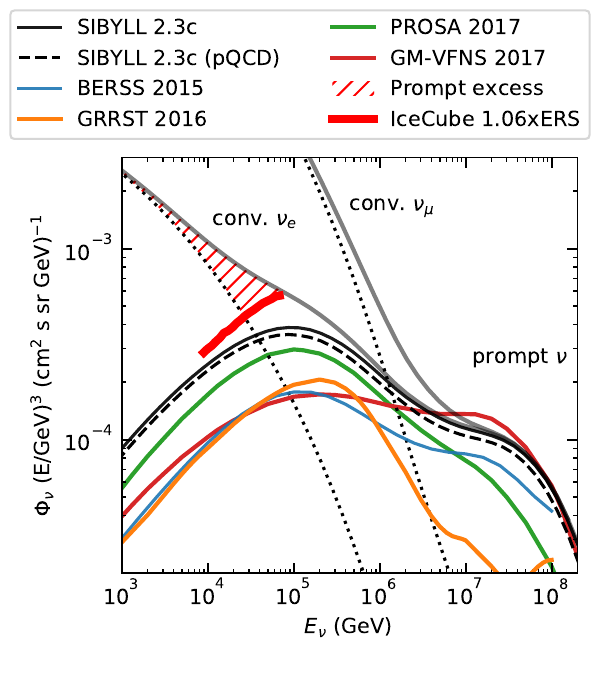}
  \caption{\label{fig:prompt_nu_flux} Prompt atmospheric muon and electron neutrino fluxes averaged over zenith angles. All fluxes are computed for the H3a cosmic ray flux model and the most differences arise from the charm cross section calculations. \sibyll{-2.3c} (pQCD) is the component attributed to the $\mathrm{perturbative}^*$ component from Figs.\ \ref{fig:sib-xl-spec} and \ref{fig:sib-charm-mvg}. The pQCD calculations BERSS \cite{Bhattacharya:2015ds}, GRRST \cite{Gauld:2015kvh,Gauld:2015yia}, PROSA \cite{Garzelli:2016xmx} and GM-VNFS \cite{Benzke:2017yjn} are computed at NLO precision under different assumptions for vertical angles. For $\theta = 0^\circ$ the shape of the black curves would coincide with the shape of the green curve. They are accompanied by large error bands which are omitted for clarity. Within the errors all models are compatible with each other. The bound by IceCube \cite{Aartsen:2016xlq} of 1.06x is expressed in units of normalization of the model from \cite{enberg_2008}, re-computed for the H3a primary flux. The red hatched patch represents the excess of the prompt flux that leads to distinct signatures in the $\nu_\mu / \nu_e$ and the vertical-to-horizontal ratio.}
\end{figure}
 
In \figu{prompt_nu_flux} the prompt lepton flux from \sibyll{-2.3c} is compared to some of the recent prompt flux calculations by \cite{Bhattacharya:2015ds,Gauld:2015yia,Garzelli:2016xmx}. An advantage of the charm model implemented in a Monte Carlo event generator is the integration into air-shower codes like CORSIKA \cite{CORSIKA_report}, where it can be used to generate exclusive air-shower or muon bundle events containing the decay products of charmed hadrons and of unflavored mesons. As discussed in Sect.~\ref{ssec:comp_with_other_charm}, the phenomenological approach to heavy flavor production in \sibyll{-2.3} yields charm cross sections comparable to other contemporary perturbative QCD calculations. The other inputs of the prompt flux computation, such as the proton-air cross section or the elasticity, have some influence, as well. All of the available models are compatible with the LHC measurements and the IceCube bound from \cite{Aartsen:2016xlq} \revise{within errors}. As expected from Figs.~\ref{fig:sib-xl-spec} and \ref{fig:sib-charm-mvg} the ``perturbative'' production from hard processes accounts for the largest fraction of the flux. The remaining contributions from diffraction, remnant excitation and fragmentation account only for a very small part and affect almost exclusively the description of the low energy data, as explained in Sect.~\ref{ssec:sib23_charm}.

The prompt electron neutrinos dominate over the conventional at much lower energies. For up-going neutrinos this transition can occur as low as a few tens of TeV. For muon neutrinos this transition happens in the PeV range due to the higher conventional component. The prompt flux is identical for both neutrino flavors, since the leptonic branching ratios of D mesons are almost equal for electron and muon flavors. Atmospheric tau neutrinos are suppressed by one order of magnitude (see \figu{hadron_contribution}). Leptons from decay of B mesons do not constitute more than 10\% of the prompt flux \cite{Martin:2003us}.

The detection of the prompt flux with instruments like IceCube through an excess of the flux is extremely challenging due to the astrophysical neutrino ``background'' that overshoots the prompt flux by almost an order of magnitude. As discussed later in Sects. \ref{ssec:neutrino_ratios} and \ref{ssec:vh_ratios}, the more sensitive observables are the flavor and the angular ratios, which would be affected by the excess of the prompt over the conventional electron neutrinos drawn as hatched area in \figu{prompt_nu_flux}. There are caveats, though. The spectral index of astrophysical neutrinos is under investigation and it is currently compatible with values between $-2$ and $-3$ \cite{Aartsen:2016xlq}. The limit by IceCube can be thought of as single power-law extrapolation of the current best fit of the through-going astrophysical muon neutrino flux at higher energies \cite{Aartsen:2016xlq}. In case future studies confirm either a broken power-law (soft at lower, hard at high energies), or, a generally soft astrophysical flux, the hatched area will shrink, making the prompt flux impossible to measure with neutrinos, at least in the standard (1:1:1) flavor scenario. The positive side effect of such a scenario is that the prompt flux would become a negligible background for neutrino astronomy. An alternative method to measure the prompt flux might involve atmospheric muons. As discussed in Sect.\ \ref{ssec:relevant_hadrons}, the prompt muon flux is partly independent of the prompt neutrino flux due to the extra component from unflavored meson decays and from electromagnetic pair production.

\subsection{Primary flux dependence of neutrino fluxes}
\label{sec:prim_flux_dependence_neutrinos}
\begin{figure*}
  \includegraphics[width=\textwidth]{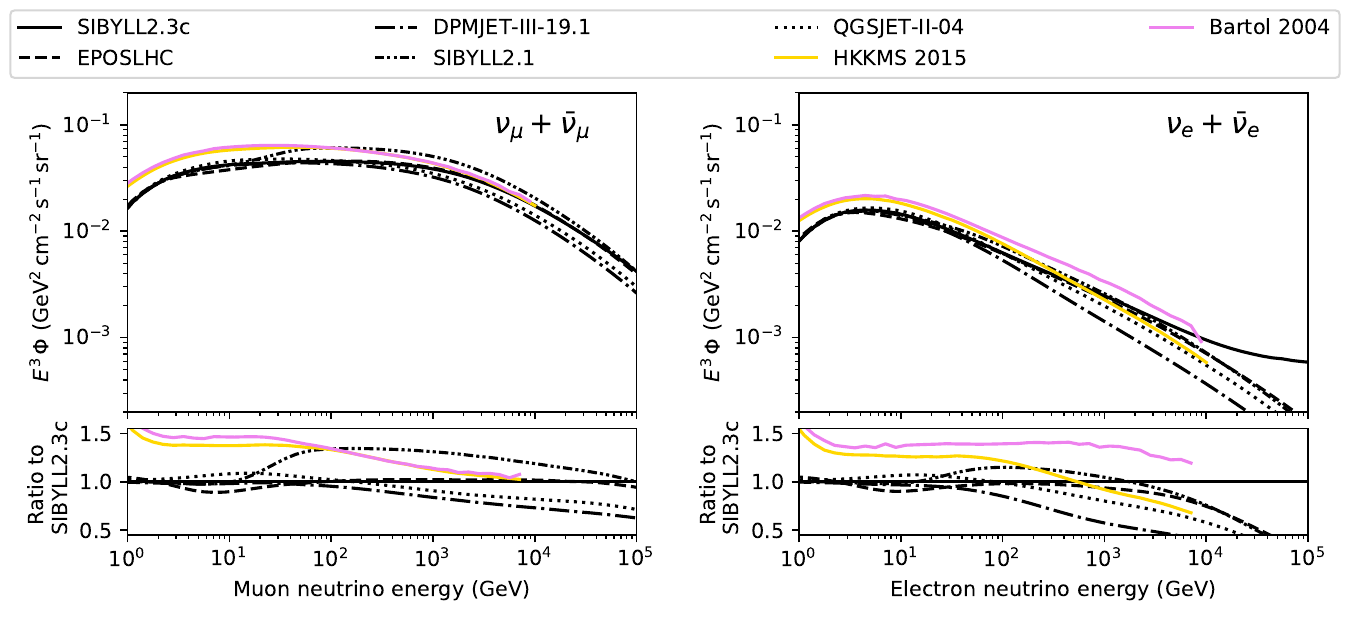}
  \caption{\label{fig:zenith_averaged_neutrinos_GSF} Atmospheric muon and neutrino fluxes averaged over the zenith angle. The Figure is similar to Fig.~\ref{fig:zenith_averaged_neutrinos}, but instead of the H3a primary flux the nucleon flux from the Global Spline Fit model (GSF) \cite{Dembinski:2017zsh} has been used. The differences in the shape and spectral index of the zenith-averaged neutrino flux are significantly reduced in the comparison between \mceq{} and the HKKMS or Bartol calculations.}
\end{figure*}
\begin{figure*}
  \includegraphics[width=\textwidth]{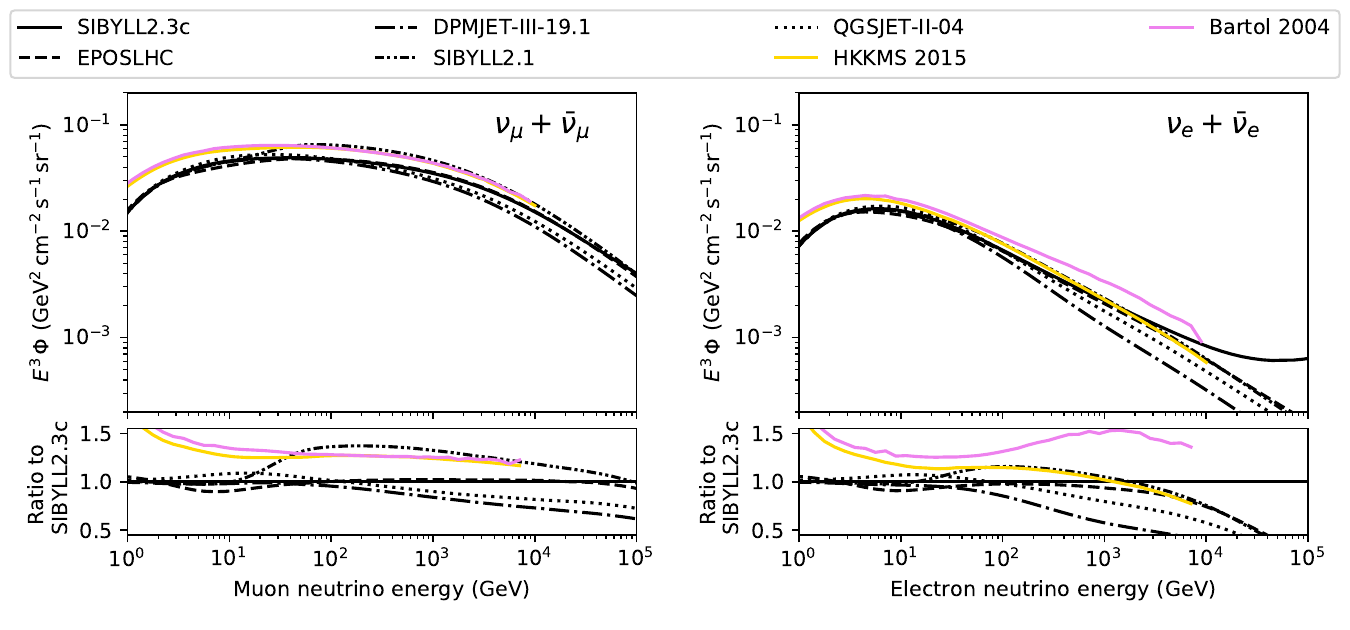}
  \caption{\label{fig:zenith_averaged_neutrinos_GH} Figure identical to Figs.~\ref{fig:zenith_averaged_neutrinos} and \ref{fig:zenith_averaged_neutrinos_GSF} but with the primary flux model from \cite{Gaisser:2002jj} that is used in the HKKMS and Bartol calculations.}   
\end{figure*}
\revise{In this section we discuss two versions of Fig.~\ref{fig:zenith_averaged_neutrinos} computed with different primary flux models that may help in understanding the origin of the differences between \mceq{} and the HKKMS or Bartol calculations.

In Fig.~\ref{fig:zenith_averaged_neutrinos_GSF}, we repeated our calculations using a recent data-driven primary flux model, the Global Spline Fit (GSF) \cite{Dembinski:2017zsh}. The agreement in the spectral index becomes better compared to Fig.~\ref{fig:zenith_averaged_neutrinos}. For energies above a few GeV, the differences between the model choices in \mceq{} and HKKMS/Bartol calculations reduce to a almost constant offset. In particular, the \qgsjet{} and \dpmjet{} lines are lower by approximately 20\% for electron and 30\% for muon neutrinos with respect to HKKMS or Bartol. 

Fig.~\ref{fig:zenith_averaged_neutrinos_GH} shows calculations performed with identical all-nucleon spectrum from \cite{Gaisser:2002jj}. Interestingly, the shape of the zenith-averaged neutrino spectrum better agrees with HKKMS using ``unmodified'' hadronic interaction models and a sophisticated primary flux model (GSF). Recall that in HKKMS the hadronic interaction model (an older version of \dpmjet{}) received energy-dependent modifications of the secondary particle yields to match inclusive muon observations \cite{Honda:2006qj}. These corrections are implemented as a series of energy dependent polylines (for details see \cite{Sanuki:2006yd}) and can explain the observed 20 -- 30\% differences. Our result (compare the lower panels of Figs.~\ref{fig:zenith_averaged_neutrinos_GSF} and \ref{fig:zenith_averaged_neutrinos_GH}) suggests that the energy dependence of the modifications to hadronic models may be not as strong as it was required for outdated primary fluxes and that the required spectral hardening, diagnosed in \cite{Honda:2006qj} from muon data, is an effect of the harder primary flux. For a potential ``calibration'' of neutrino fluxes with muon data one has to be careful in accurately treating the degeneracy between the primary flux and the hadronic model. Ongoing studies suggest that good fits to inclusive muon data can be achieved by allowing for energy-independent constant scaling factors for pions and kaons \cite{Yanez:2019bnw}.}

\subsection{Angular dependence}
The dependence of fluxes and flavor ratios on zenith directions is a crucial observable for the measurement of fundamental neutrino properties. In atmospheric neutrino oscillation studies with volumetric detectors, the zenith angle defines the baseline distance over which neutrinos can oscillate. This technique has been applied to obtain evidence for neutrino oscillations with the Super-Kamiokande experiment \cite{Kajita:2016vhj}. Recent experiments, such as IceCube/DeepCore, evolved this method and make use of larger detector volumes and different energy bands to determine oscillation parameters to a precision competing with dedicated accelerator setups \cite{Aartsen:2017nmd}. At much higher energies, the pattern in the zenith-energy plane gives access to physics beyond standard model (BSM) scenarios \cite{TheIceCube:2016oqi}. Future experiments \cite{Aartsen:2014oha,Adrian-Martinez:2016fdl} will increasingly rely on accurate predictions of angular distributions and require access to the underlying uncertainties of the physical models. In the following sections we scrutinize the calculations obtained from the combination \mceq{} + \sibyll{-2.3c} against the available reference models.

\subsubsection{Neutrino fluxes}

\begin{figure*}
  \includegraphics[width=\textwidth]{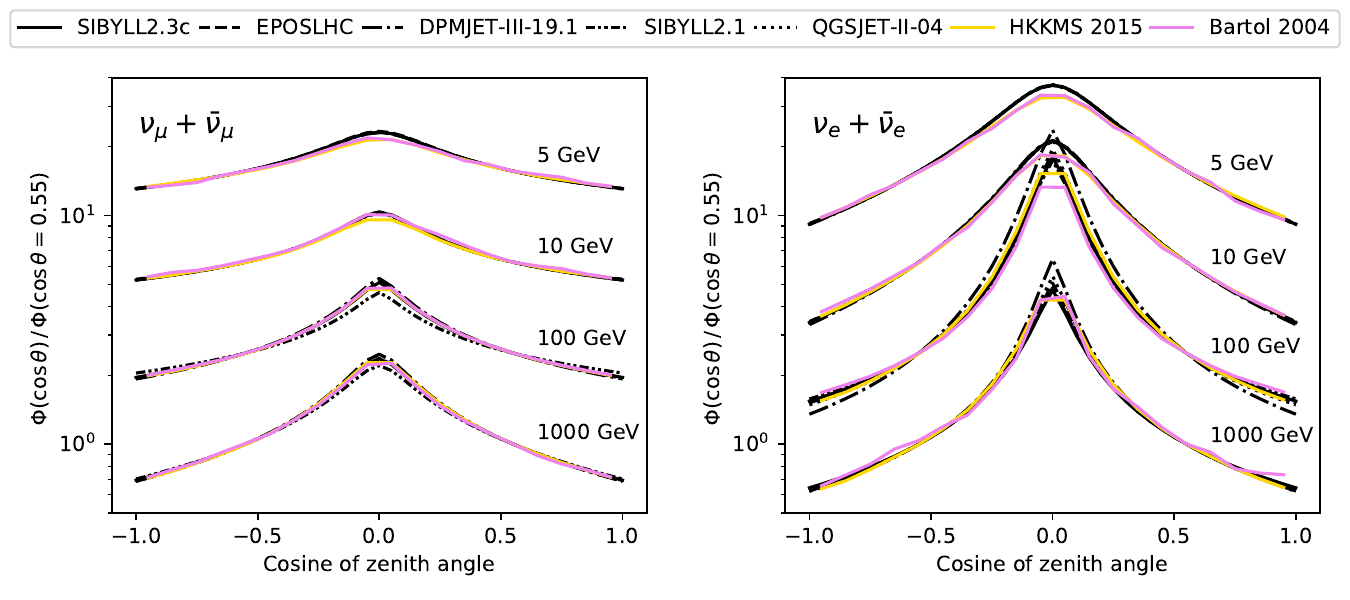}
  \includegraphics[width=0.98\textwidth]{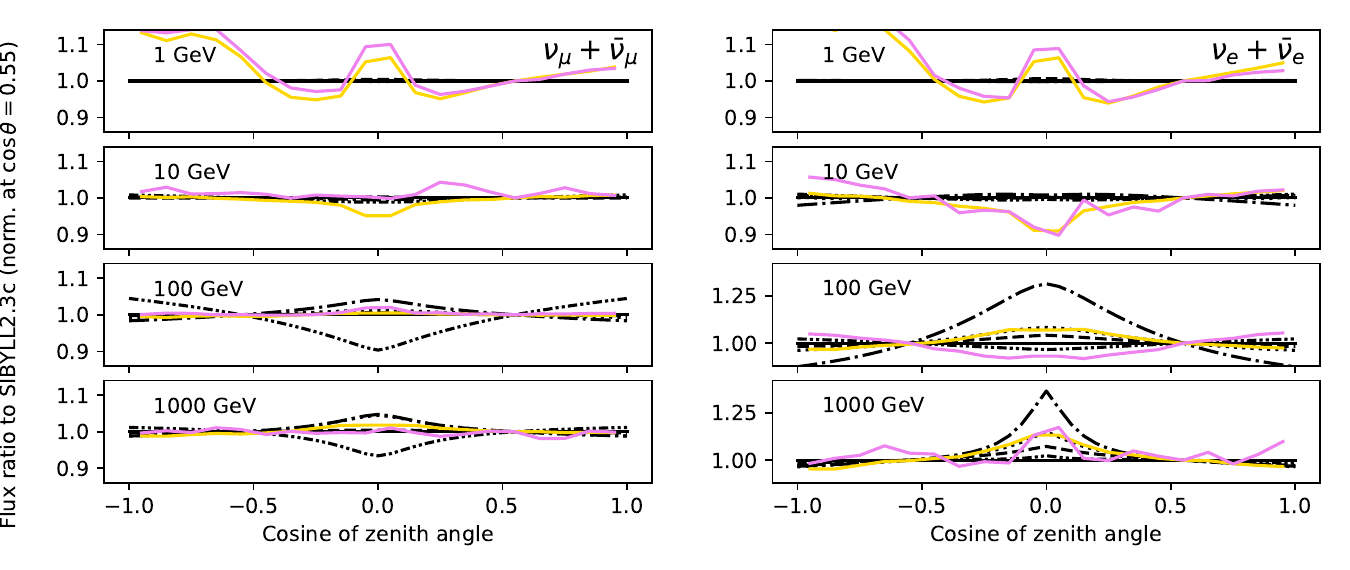}
  \caption{\label{fig:angular_neutrino_distr} Azimuth-averaged zenith distributions at fixed neutrino energies. In the upper panels, each individual curve is normalized to one at $\cos{\theta} = 0.55$ and offset through a multiplication with $2.5$. The lower panels show the model ratios, normalized at $\cos{\theta} = 0.55$ to the \sibyll{-2.3c} value.}
\end{figure*}

The zenith angle distributions (see \figu{angular_neutrino_distr}) of fluxes agree between the calculations within a few to ten percent with some exceptions. Excluding \sibyll{-2.1}, the dependence on the hadronic interaction model is weak. Muon neutrino fluxes are particularly sensitive to the K$^+$ abundance at high energies, since muon neutrinos originate primarily from decay of charged kaons in the TeV-PeV range.  

The differences between the Monte Carlo calculations (HKKMS and Bartol) and \mceq{} do \revise{originate from several contributing factors (ordered by impact): hadronic interactions, primary fluxes, geometry and calculation method.} For muon neutrinos deviations are generally small and at $5\,$GeV, \revise{some of the effects that \mceq{} does not include, for instance the geomagnetic cutoff or the onset of 3D effects, lead to larger discrepancies.} Electron neutrinos show larger differences at the horizon (note the increased y-axis scale in the lowest right panels). \revise{This region is in particular sensitive to kaon yields and to differences in the inelastic kaon-air cross section that is not well constrained at high energies. The \dpmjet{} calculations show deviations from the predictions of the other models for high-energy electron neutrinos. This is related to a a significantly lower productions of strange hadrons, which is likely related to the updated parton distribution functions \cite{Fedynitch:2015kcn}.}

\subsubsection{Muon fluxes}
\begin{figure*}
  \includegraphics[width=\textwidth]{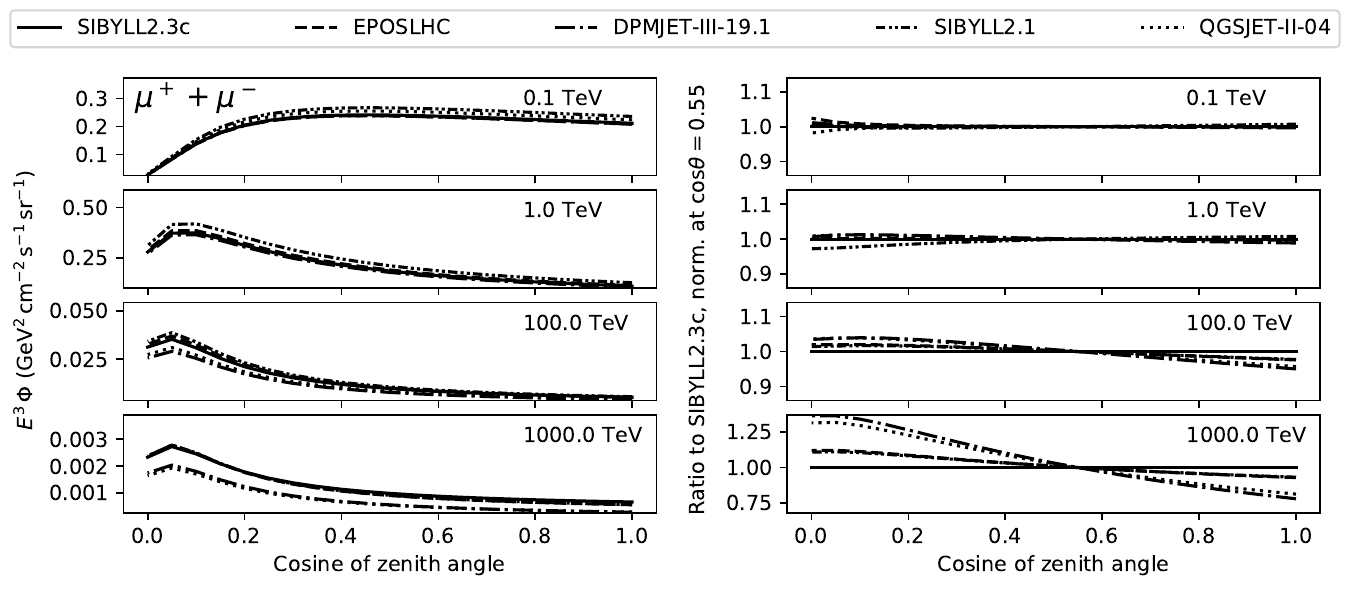}
  \caption{\label{fig:angular_muon_distr} Azimuth-averaged zenith distributions of the total (conv. + prompt) atmospheric muon flux at high energies. The angle $\cos{\theta} = 0$ corresponds to horizontal directions and $\cos{\theta} = 1$ to vertical down-going.  The left panels show the muon fluxes computed with different hadronic models. Their ratios to \sibyll{-2.3c} are located on the right hand side and normalized to one at $\cos{\theta} = 0.55$. Note the increased scale of the lowest right panel.} 
\end{figure*}

The impact of the hadronic model on the angular distribution of muon fluxes is demonstrated in \figu{angular_muon_distr}. At lower energies, where the muon flux is dominated almost exclusively by the pion component, the models behave similarly. However, at very high energies the angular distributions do not match. This is related to the prompt muon flux from unflavored and charmed meson decays (compare with \figu{hadron_contribution}). In addition, some contribution from muon pair production can be expected that is not accounted for in the present calculation. We checked that the angular distributions at high energy exactly match for \revise{purely} conventional fluxes. The previous attempts by the IceCube Collaboration to measure the atmospheric muon fluxes were negatively impacted by a mismatch in the zenith distribution \cite{Abbasi:2012kza, Aartsen:2015nss}. While a part of the problems may originate from experimental uncertainties, it would be worth to revisit these measurements with the post-LHC interaction models. As outlined below and in Sect.~\ref{ssec:prompt_flux_discussion} for neutrinos, the presence of a prompt component affects in particular the angular distributions making them flatter than the conventional-only scenario. 

\subsubsection{Neutrino ratios}
\label{ssec:neutrino_ratios}
\begin{figure*}
  \includegraphics[width=\textwidth]{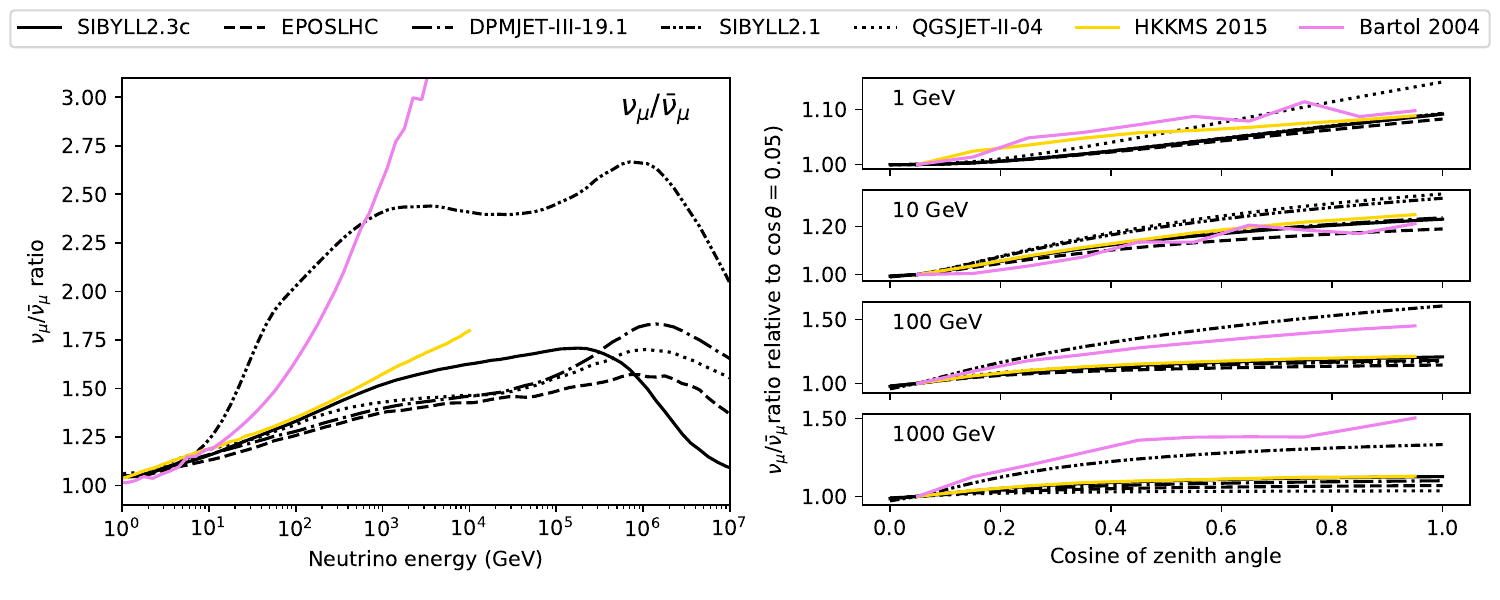}
  \includegraphics[width=\textwidth]{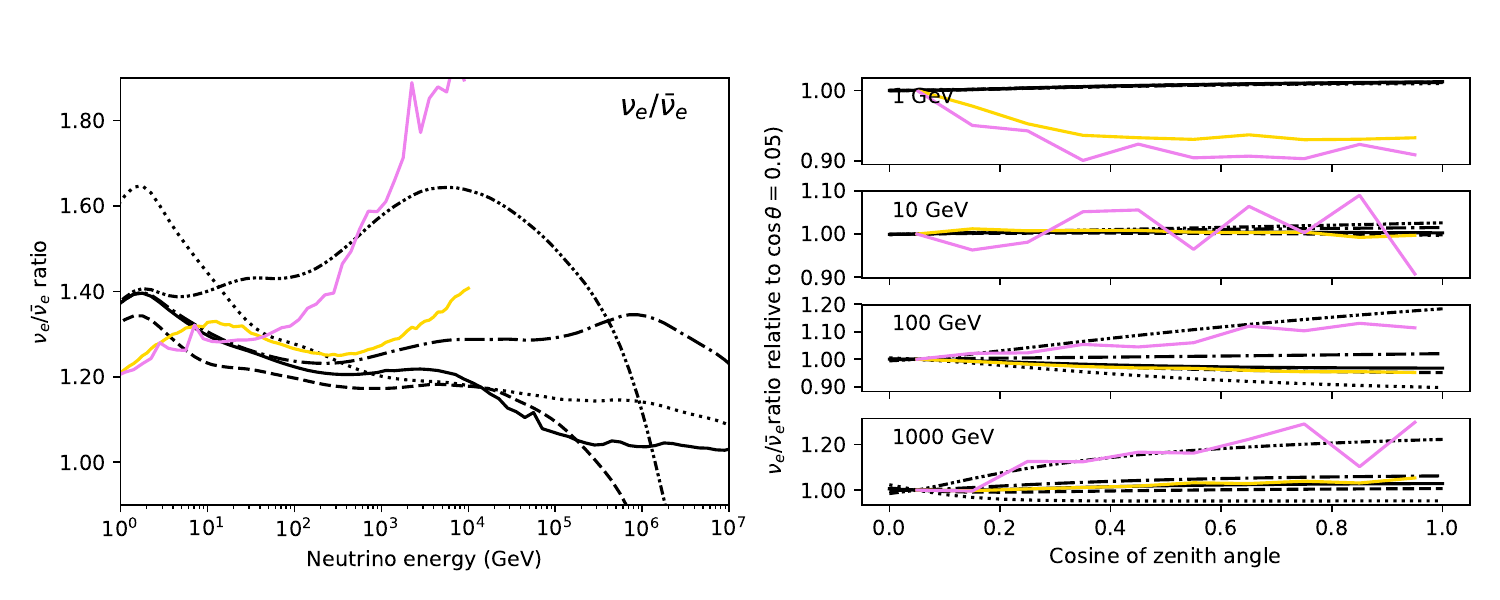}
  \includegraphics[width=\textwidth]{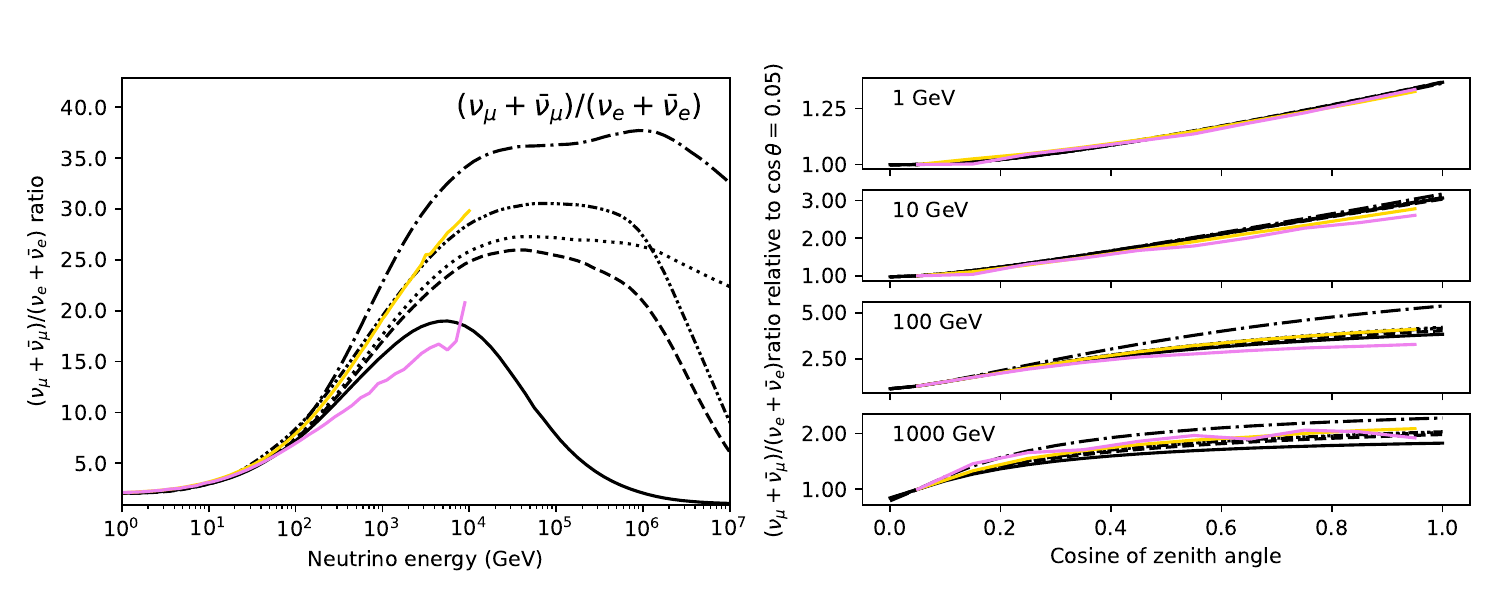}
  \caption{\label{fig:neutrino_ratios} Neutrino-antineutrino and flavor ratios. The left panels show zenith-averaged distributions. The right panels demonstrate the relative zenith angle dependence of each ratio for a fixed neutrino energy. These curves are normalized such that at $\cos{\theta} = 0.05$ the value is one. Note that the axes scales for the right panels were allowed to float.}
\end{figure*}
The \figu{neutrino_ratios} gives a detailed overview on the behavior of the neutrino ratios. Ratios trace particularly well hadronic interactions, since the dependence on cosmic ray flux mostly cancels out. As illustrated before in \figu{angular_distribution} the angular spectrum encodes the presence of different hadronic species and it is instructive to involve that figure in the present discussion. 

The muon neutrino/anti-neutrino ratio at higher energies in the upper panels separates the calculations, or more precisely the hadronic models of these calculations, into two classes. Those with a very enhanced forward K$^+$ production and those without a particular emphasis on this channel. The Bartol and the \sibyll{-2.1} curves show a similar trend related to the high abundance of K$^+$, while the \eposlhc{} model has contrary trend albeit less significant.

The muon-neutrino/anti-neutrino ratio inferred from the energy-dependence of the muon charge ratio in \cite{Gaisser:2012em} is closer to the \sibyll{-2.1} value, rising from 1.5 at 10 GeV to 2.2 in the TeV range and above.  This is a consequence of the fact that,
because of the two-body decay kinematics, most muon neutrinos come from
decays of charged kaons rather than pions above 100~GeV.  Fitting the increase in the measured muon charge ratio, after accounting for the neutron fraction in the primary cosmic-ray beam, normalizes the kaon contribution to the muon flux.  The resulting muon
charge ratio obtained in this way increases from 1.28 at 10 GeV to 1.41 at 10 TeV, somewhat higher than the corresponding ratio from \sibyll{-2.3c} in \figu{vertical_muon_fluxes}.  The corresponding muon-neutrino/anti-neutrino ratio is amplified by the kinematic effect, so that its difference from \sibyll{-2.3c} in \figu{neutrino_ratios} is greater.

The electron neutrino/anti-neutrino ratio (middle panels of \figu{neutrino_ratios}) processes similar discrepancies among the models. The abundance of K$^+$ dictates the ratio at energies above a hundred GeV. The up-turn of the HKKMS calculation above a TeV might be a relic of the corrections applied to fit the muon charge ratio, and this behavior can indeed be more realistic than the flatter distribution predicted by the interaction models in \mceq{}. During the development of \sibyll{-2.3c}, we aimed to have an accurate microscopical description for the muon charge ratio, but despite a significant improvement the result is not perfect. On the other hand, our extrapolations are based on a self-consistent model and are, therefore, better suited for extrapolations to higher energies.

The flavor ratio in the lower panels is less sensitive to variations in the secondary hadron production. At low energies, where electron neutrinos originate from decaying muons, the calculations agree well since muon neutrinos and muons are both coming from pion decay. In this case the muon to electron neutrino ratio is fixed by the decay kinematics of muons. At higher energies this ratio depends more on the hadronic model since electron neutrinos have an independent production channel through decays of short and long states of neutral kaons.

The prompt flux in \sibyll{-2.3c} gives rise to a ratio of one at the highest energies, since the branching ratios of charmed mesons are similar for muon and electron neutrino flavors. This impacts the expected track-cascade (muon-line/electron-like events) ratio in neutrino telescopes. As the lower left panel clearly demonstrates, the flavor ratio moves towards one since the prompt muon and electron neutrino fluxes are equal. The deviation from a conventional-only hypothesis emerges at energies as low as $10\,$TeV and it is not very dependent on the hadronic model. Close to $100\,$TeV this difference is striking and must yield a sensitive observable in next generation neutrino telescopes (with larger effective areas for the cascade channel). One caveat is the presence of the astrophysical flux that is currently compatible with the same muon to electron neutrino ratio of one \cite{Aartsen:2015ivb}.

\subsubsection{Vertical-to-horizontal ratio}
\label{ssec:vh_ratios}
\begin{figure*}
  \includegraphics[width=\textwidth]{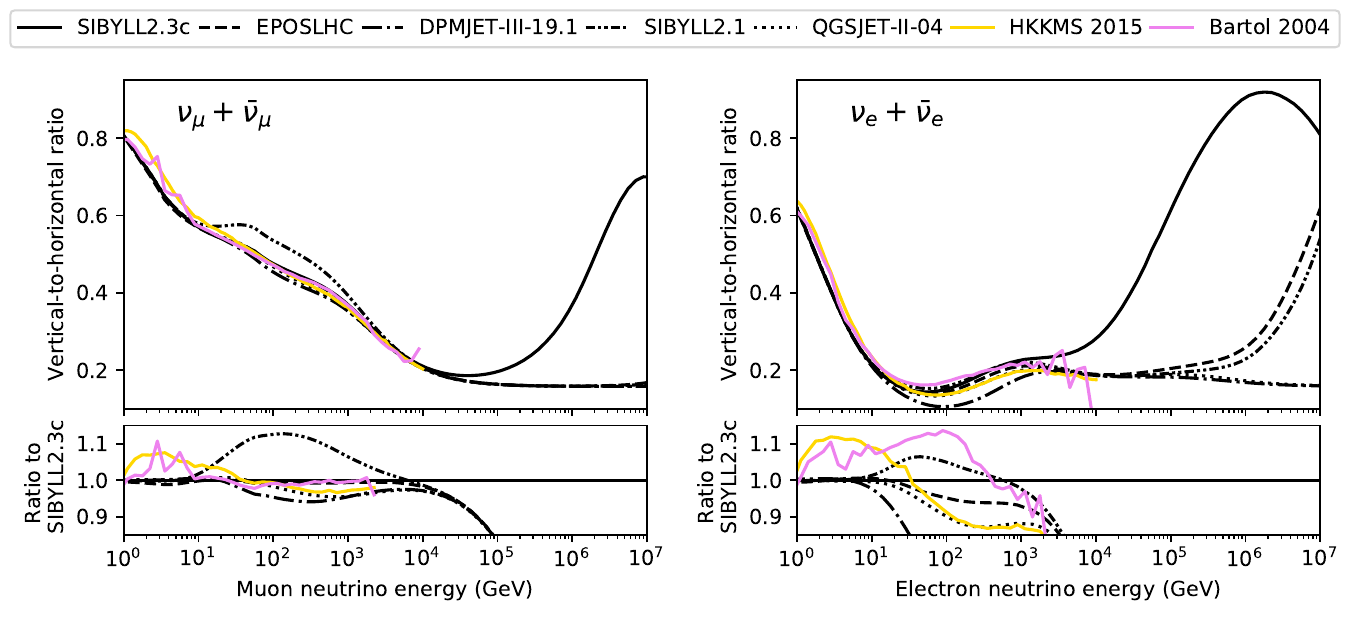}
  \caption{\label{fig:down_horizontal} Down to horizontal ratio. It is defined as the ratio of fluxes integrated in the angular bin $\cos{\theta} < 0.4$ around the two extreme directions. The comparison is shown down to the lowest energies, since it is stongly impacted by 3D calculations at energies below a few GeV. Large differences at high energy originate from the presence of prompt neutrinos in the \sibyll{-2.3c} calculations.}
\end{figure*}
For completeness, we discuss the vertical-to-horizontal ratio (this calculation is up-down symmetric) using the same definition as in \cite{Barr:2004br} (within $\cos{\theta} < 0.4$ around the vertical and horizontal directions). This ratio is sensitive to the differences between 3D and 1D calculations and it is shown down to the lowest energies for both reference calculations and \mceq{} in \figu{down_horizontal}. 

For muon neutrinos the ratios agree within a few percent between \mceq{} and the Bartol calculation, which switches to full-3D at $10\,$GeV. At energies below $30\,$GeV (where the HKKMS calculations switches to 3D) there is an observable shift between \mceq{} and HKKMS that stays below 10\%. The old \sibyll{-2.1} notably disagrees at medium energies due to charged kaons and the resulting error in the production height that is different for the muon neutrinos from pion decays. Apart from this, the dependence on the details of the hadronic model is weak. 

In electron neutrinos the differences among the hadronic models are much larger, reflecting the higher uncertainty in the production mechanisms of charged and neutral kaons that are still fundamentally not understood in particle physics, unfortunately. Due to the lack of access to the pion and kaon components of the reference calculations the exact origin of these different behaviors is difficult to trace. 

The angular distribution of electron neutrinos at high energies is significantly affected by the prompt flux. The right panel shows that (even when using large zenith bins of $\cos{\theta} < 0.4$), the expected flux of the vertical-to-horizontal ratio is a very sensitive observable. As mentioned above, the detection of the prompt neutrino flux with a future neutrino telescope in the cascade channel would not require excessively high energies that otherwise would impact the statistical errors. The two signatures, the angular distribution of cascades and the above mentioned track/cascade ratio, are sensitive to the flux excess in the (upper) triangle enclosed by the dotted black, the solid gray and the red upper limit line in \figu{prompt_nu_flux}. A successful determination of the prompt flux will remain a tough experimental challenge and almost certainly require an upgraded detector and a better characterization of the astrophysical flux.

\section{Summary}

This work is about the connection between hadronic interactions and the inclusive fluxes of muons and neutrinos in the Earth's atmosphere. The numerical solution of the transport equations with \mceq{} provides a study of this connection up to the highest energies at very high precision, being essentially free from statistical simulation uncertainties. 

We characterized the distributions of cosmic ray energies that can produce leptons at certain energies at ground, taking into account the effect of a non-power-law primary spectrum affected by the knee and the ankle of cosmic rays. Essentially the entire energy range is accessible at particle colliders, but in the center of mass frame. However, the collider kinematics and the lack of proton--light-ion collisions imposes limitations on the applicability of these data to our case. We identify all types of hadrons that contribute to the inclusive fluxes and how the interplay between the production cross section and decay time impacts the zenith distributions. 

The atmospheric muons, which can be measured with high precision, behave in several ways differently from the atmospheric neutrinos. The latter can only be assessed inclusively, \ie{} integrated over the primary cosmic ray energy, while muons can be studied also in exclusive events like air-showers or muon bundles. Of particular importance at high energies is the prompt flux that requires a model for the production of heavy flavor hadrons. It is an integral part of the new \sibyll{-2.3c} model and it is discussed in greater detail. The prompt flux of muons is different from the prompt neutrino flux and we characterized these contributions from unflavored mesons. The muon charge ratio is known to be sensitive to the forward particle production. To constrain the forward pion and kaon production, one needs to account both for the primary spectrum of nucleons (separating
protons and neutrons) and for the shape of the particles' longitudinal energy spectra.
Because the fluxes in this work are calculated with a single model of the primary spectrum, a full account of the lepton ratios needs further work. 

We show that the longitudinal spectra at high $x_{\rm F}$ or $\xl{} > 0.2$ are paramount for inclusive fluxes and introduce a new non-perturbative process in the \sibyll{-2.3c} interaction model that gives additional degrees of freedom to better reproduce the leading particle effect, a forward flavor asymmetry that originates from the high momentum fraction carried by the valence quarks of the projectile. This mechanism in the new version of \sibyll{} is the remnant excitation model, in which the valence quarks are separated from the sea partons and allowed to fragment independently. The remnant model gives a significantly improved description of fixed-target data, resulting in a better, albeit not yet perfect, prediction of the muon charge ratio. We made sure that the new version is approximately compatible with Feynman scaling in the forward phase space as observed in the data and which is a central element of the Dual Parton Model.

The new model for the production of heavy flavors in \sibyll{} is based on the family relation between strange and charmed quarks. At different stages of the event generation, charmed quarks can be produced through the replacement $s \to c$ with certain transition probabilities that have been determined by comparison to a large variety of data sets on production of charmed hadrons. The large mass of the charm quark is beyond the non-perturbative scale, and therefore, charm is predominantly produced in perturbative (hard) processes. We find that the augmented minijet model is sufficient to describe the total yields of charm at all accessible energies from fixed-target experiments up to the LHC. The differential ($p_T$) spectra are tolerably described, but show some tension towards forward LHCb rapidities. We find that the hard component is insufficient to describe charm production at fixed-target energies at forward $x_{\rm F}$, which requires accounting for processes such as associated production. 
At high energies, where the charm production is relevant for atmospheric neutrino fluxes, the dominant contribution comes from the perturbative component and this scenario is in agreement with other contemporary calculations of the prompt flux. In extensive comparisons with NLO calculations, we generally find an agreement between our simplified approach and the more sophisticated methods within their uncertainties.

Finally, we benchmark the combination of \sibyll{-2.3c} and \mceq{} against other reference calculations including full three-dimensional Monte Carlo calculations. For \mceq{} we also employ other interaction models to better disentangle the impact of hadronic interactions from the cascade physics or the calculation method. We generally find a good agreement when using our methods that require a tiny fraction of computational time. Most differences arise from the modeling of forward kaon production that is the most uncertain component in the prediction of atmospheric fluxes and flavor ratios. However, there are additional features in the angular distributions close to the horizon that do not seem to come from differences in hadronic interactions and more likely stem from the calculation methods. The \sibyll{-2.1} model is shown to overproduce K$^+$ with a notable impact on many observables. Therefore we discourage users from employing this model in future calculations of inclusive fluxes and use instead the new version or one of the other interaction models. The new charm model predicts a prompt flux that is somewhat higher than the central expectations of the other current models (within errors) and it is also compatible with the experimental limit by IceCube. We discuss prospects for measuring prompt neutrinos at current or future neutrino telescopes and outline a number of distinct signatures that can be assessed through the cascade channel at moderate energies between 10 - $100\,$TeV. The impacted variables are the muon-to-electron neutrino ratio and the vertical-to-horizontal ratio that are both sensitive to the angular distribution and the track-to-cascade ratio in volumetric detectors.

\acknowledgments
For their steady interest, the inspiring discussions and for the early adoption of our codes, we would like to thank our colleagues from the IceCube Collaboration, in particular Dennis Soldin, Summer Blot, Jakob van Santen, Jason Koskinen and Tyce de Young. We also thank Maria Vittoria Garzelli for valuable discussions and for in-depth comparisons between the charm models.  

This project has received funding from the European Research Council (ERC) under the European Unions Horizon 2020 research and innovation programme (Grant No. 646623). Work of T. Gaisser and T. Stanev is supported in part by grants from the U.S. Department of Energy (DE-SC0013880) and the U.S. National Science Foundation (PHY 1505990). 
The work of F.\ Riehn is supported in part by the KIT graduate school KSETA, in part by the German Ministry of Education and Research (BMBF), grant No.\ 05A14VK1, in part by the Helmholtz Alliance for Astroparticle Physics (HAP), which is funded by the Initiative and Networking Fund of the Helmholtz Association and in part by OE - Portugal, FCT, I.\ P.~, under project CERN/FIS-PAR/0023/2017 and OE - Portugal, FCT, I.\ P.~, under project IF/00820/2014/CP1248/CT0001.

The authors would like to express a special thanks to the Mainz Institute for Theoretical Physics (MITP) of the DFG Cluster of Excellence PRISMA+ (Project ID 39083149), for its hospitality and support. AF completed parts of this work as JSPS International Research Fellow. 


%

\end{document}